\documentclass{aa}  
\usepackage{textgreek}
\usepackage{graphicx}
\usepackage{comment}
\usepackage{mathtools}
\usepackage{amsmath}
\usepackage{fancyref}
\usepackage{longtable}
\usepackage{rotating}
\usepackage{pdflscape}
\usepackage{enumerate}
\usepackage{xtab,booktabs}
\usepackage{textcomp}
\usepackage[citecolor=blue, urlcolor=blue, linkcolor=blue, colorlinks=true]{hyperref}
\usepackage{txfonts}
\usepackage{soul}
\usepackage{natbib}
\usepackage{subcaption}
\usepackage{bm}

\newcommand{\kms}{$\mathrm{km\,s^{-1}}$}

\newcommand{\angstrom}{\textup{\AA}}

\usepackage{afterpage}
\usepackage{caption}
\usepackage{float}
\usepackage{lipsum}
\usepackage{ulem}

\begin{document}

   \title{The HD 98800 quadruple pre-main sequence system}
   \subtitle{Towards full orbital characterisation using long-baseline infrared interferometry\thanks{Based on observations made with ESO telescopes at Paranal under the program IDs 60.A-9131, 0104.C-0556, 105.20JX.}}

  \author{S. Z\'u\~niga-Fern\'andez 
          \inst{1, 2, 3} \thanks{sebastian.zuniga@postgrado.uv.cl}
          \and{J. Olofsson}
          \inst{3, 1}
          \and A. Bayo 
          \inst{3, 1}
          \and{X. Haubois}
          \inst{2}
          \and{J. M. Corral-Santana}
          \inst{2}
          \and{A. Lopera-Mej\'ia}
          \inst{3}
          \and{M.\,P.\,\,Ronco}
          \inst{1,4}
          \and{A. Tokovinin}
          \inst{5}
          \and{A.\,Gallenne}
          \inst{7,8,9}
          \and{G. M. Kennedy}
          \inst{6}
          \and{J.-P. Berger}
          \inst{10}
          }
   \institute{
   N\'ucleo Milenio de Formaci\'on Planetaria (NPF), Valpara\'iso, Chile
   \and
    European Southern Observatory, Alonso de C\'ordova 3107, Vitacura, Casilla 19001, Santiago de Chile, Chile
   \and
    Instituto de F\'isica y Astronom\'ia, Facultad de Ciencias, Universidad de Valpara\'iso, Av. Gran Breta\~na 1111, Valpara\'iso, Chile
    \and
    Instituto de Astrof\'{\i}sica - Pontificia Universidad Cat\'olica de Chile, Av. Vicu\~na Mackenna 4860, Macul - Santiago, 8970117 , Chile
    \and
    Cerro Tololo Interamerican Observatory | NSF's NOIRLab, La Serena, Casilla 603, Chile
    \and
    Department of Physics, University of Warwick, Gibbet Hill Road, Coventry, CV4 7AL, UK
    \and
    Nicolaus Copernicus Astronomical Centre, Polish Academy of Sciences, Bartycka 18, 00-716 Warszawa, Poland
    \and
    Universidad de Concepción, Departamento de Astronomía, Casilla 160-C, Concepción, Chile
    \and
    Unidad Mixta Internacional Franco-Chilena de Astronomía (CNRS UMI 3386), Departamento de Astronomía, Universidad de Chile, Camino al Observatorio 1515, Las Condes, Santiago, Chile
    \and
    Université Grenoble Alpes, CNRS, IPAG, F-38000 Grenoble, France
    }
    \date{Received; Accepted}

 
  \abstract
  {HD 98800 is a young ($\sim10$ Myr old) and nearby ($\sim45$ pc) quadruple system, composed of two spectroscopic binaries orbiting around each other (AaAb and BaBb), with a gas-rich disk in polar configuration around BaBb. While the orbital parameters of BaBb and AB are relatively well constrained, this is not the case for AaAb. A full characterisation of this quadruple system can provide insights on the formation of such a complex system.}
    {The goal of this work is to determine the orbit of the AaAb subsystem and refine the orbital solution of BaBb using multi-epoch interferometric observations with the Very Large Telescope Interferometer (VLTI)/PIONIER and radial velocities.}
    {The PIONIER observations provide relative astrometric positions and flux ratios for both AaAa and BaBb subsystems. Combining the astrometric points with radial velocity measurements, we determine the orbital parameters of both subsystems.}
    {We refined the orbital solution of BaBb and derived, for the first time, the full orbital solution of AaAb. We confirmed the polar configuration of the circumbinary disk around BaBb. From our solutions, we also inferred the dynamical masses of AaAb ($M_{Aa} = 0.93 \pm 0.09$ and $M_{Ab} = 0.29 \pm 0.02$\,M$_{\sun}$). We also revisited the parameters of the AB outer orbit.}
    {The orbital parameters are relevant to test the long-term stability of the system and to evaluate possible formation scenarios of HD\,98800. Using the N-body simulation, we show that the system should be dynamically stable over thousands of orbital periods and that it made preliminary predictions for the transit of the disk in front of AaAb which is estimated to start around 2026. We discuss the lack of a disk around AaAb, which can be explained by the larger X-ray luminosity of AaAb, promoting faster photo-evaporation of the disk. High-resolution infrared spectroscopic observations would provide radial velocities of Aa and Ab (blended lines in contemporary observations), which would allow us to calculate the dynamical masses of Aa and Ab independently of the parallax of BaBb.  Further monitoring  of  other  hierarchical systems will  improve  our understanding  of  the formation and dynamical evolution of these kinds of systems.}


   \maketitle
%
\section{Introduction}


Solar-type multiple systems are at least as common as individual stars: the fraction of triple-star systems was found to be $8\pm1\%$, and it drops to $3\pm1\%$ for higher-multiplicity systems \citep{Raghavan2010}. Similarly, observations of F and G stars within 67 pc of the Sun \citep{Tokovinin2014a,Tokovinin2014b} show that
$\approx10\%$ of all stellar systems are triple and
 $\approx4\%$ are quadruple. The high-order multiplicity fraction increases with stellar mass \citep{Duchene2013}.  Multiple star systems with $n>2$ are nearly always hierarchical, meaning that they can decompose into binary or single sub-systems based on their relative separations (e.g. two close binaries that orbit each other with a wide separation). A hierarchical system can have many distinct configurations.  For instance, quadruple systems can have two possible configurations. A triple system orbited by a distant fourth companion corresponds to the 3+1 configuration. The 2+2 configuration consists in two close binaries orbiting around each other. The 2+2 configuration seems to be $\sim4$ times more frequent than the 3+1 configuration for solar-type stars \citep{Tokovinin2014b}. The orbital parameters in hierarchical systems could provide additional information about their formation history. It is expected that different formation processes, such as core fragmentation, disk instability, dynamical interactions, or a combination of different formation channels, leave imprints on the mass ratio, periods, eccentricities, and mutual orbit inclination of hierarchical systems \citep{Kozai1962, LIDOV1962,Whitworth2001,Sterzik2002,Lee2019b,TokovininMoe2020}. In the last decades, observational and theoretical efforts have led to a better understanding of the formation and dynamical stability of such multiple systems \citep{Kizelev1980,Tokovinin2006,Eggleton2009,Tokovinin2018b,Hamers2021}.

Wide binaries show a strong preference to be in hierarchical systems in low density young associations \citep{Elliott2016a,Elliott2016b} and star-forming regions \citep{Joncour2017}. The fact that this relation is not seen in the same proportion in denser environments or systems in the field suggests that this could be the result of dynamical processing or the unfolding of hierarchical systems \citep{Sterzik2002,Reipurth2012}. In that regard, characterising young (1-100 Myr), hierarchical systems helps to observe their early evolution. The formation channels of hierarchical systems cannot be easily determined by only characterising field stars, where billions of years of dynamical evolution may have erased their formation history. Consequently, the study of young (1-100 Myr) hierarchical systems is an important step to better understand their formation pathway. Large-scale surveys provide crucial information that helps to discover such multiple systems, but they are not well suited to finely constrain their orbital architecture. We need high-precision astrometry and radial velocity (RV) follow-up observations of the identified hierarchical systems to accurately constrain parameters of their inner and outer orbits.

\begin{figure}
   \centering
   \includegraphics[width=0.25\textwidth]{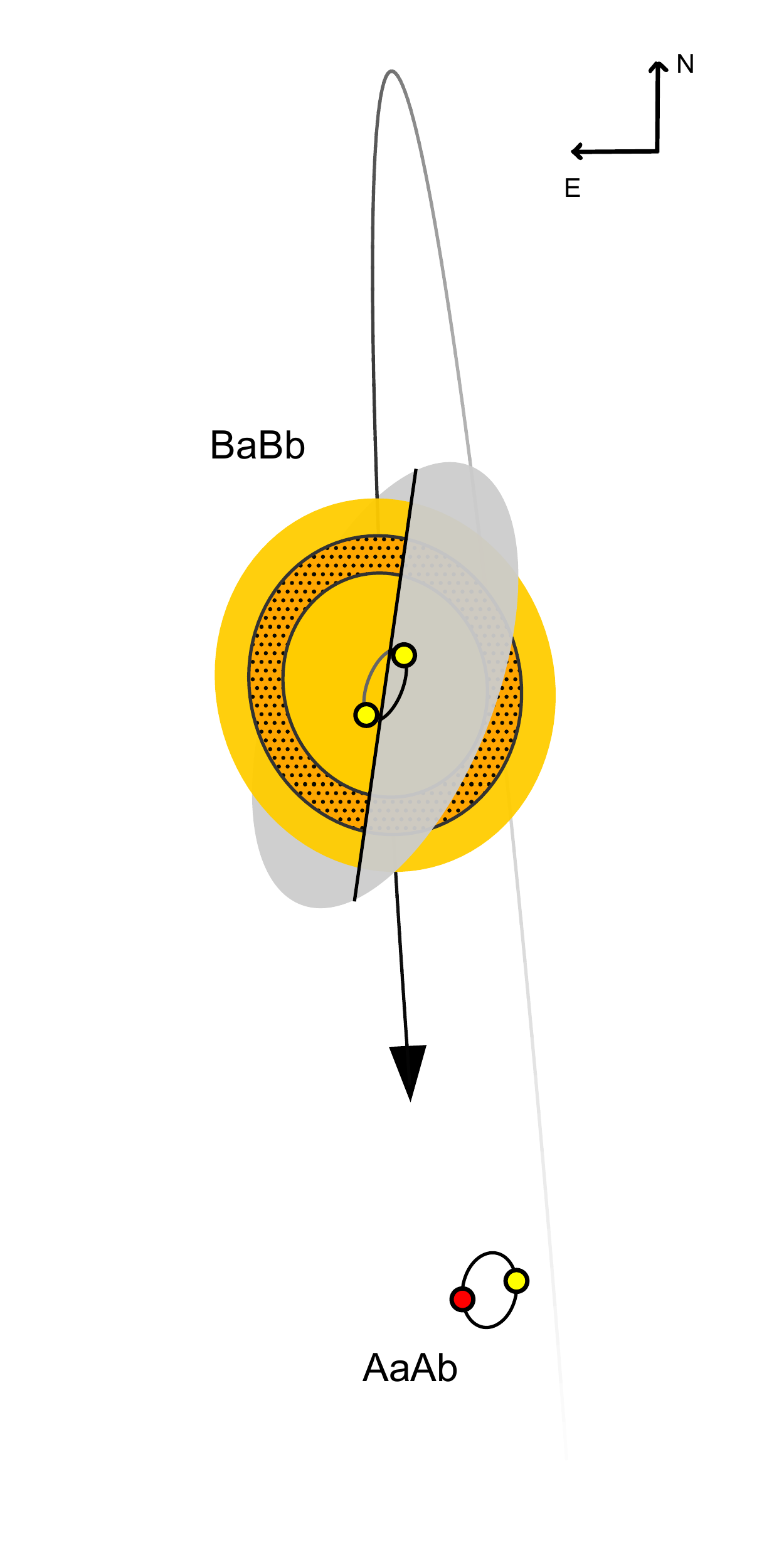}
   \caption{Schematic view of HD\,98800 orbital configuration. The BaBb subsystem hosts a circumbinary disk in polar configuration and is orbiting around the AaAb binary in a highly inclined orbit with a semi-major axis of $\approx$ 45 au.} 
   \label{Fig:system_config}
    \end{figure}

The HD\,98800 is a well-known hierarchical quadruple star system, and a member of the  10-Myr old TW Hydrae association \citep{Torres2008}. Located at $44.9\,\pm\,4.6$ pc from Earth according to the latest reduction of Hipparcos data\footnote{\tiny There is no reliable Gaia eDR3 parallax for HD 98800 \citep{GaiaMission2016,GaiaEDR32021}.} \citep{Hipparcos2007}, corresponding to a parallax of $22.27\,\pm\,2.32$ mas, it consists of two pairs of spectroscopic binaries (hereafter, AaAb and BaBb, see Fig \ref{Fig:system_config}). Both binaries orbit each other with a semi-major axis of $\approx 45$\,au \citep{Tokovinin2014c}. The AaAb system is a single-lined spectroscopic binary (SB1) with a period of 262 days \citep{Torres1995}. The mass of the Aa was estimated from pre-main sequence evolutionary models as $1.1\,\pm\,0.1\,$ M$_{\sun}$ \citep{Prato2001}. The BaBb subsystem is a double-lined spectroscopic binary (SB2) with a period of $315$ days, the \textsf{astrometric} orbital solution of this binary was first presented in \cite{Boden2005} using five Keck Interferometer (KI) epochs combined with Hubble Space Telescope astrometry, and available RV observations. From this orbital solution, \cite{Boden2005} estimated a parallax of $23.7\,\pm\,2.6$ mas, and dynamical masses for Ba and Bb of $0.699\,\pm\,0.064$ and $0.582\,\pm\,0.051$ $M_{\odot}$, respectively. The BaBb pair also harbours a bright circumbinary protoplanetary disk \citep{Skinner1992,Zuckerman1993}, and ALMA observations revealed that the disk and the binary orbital planes are perpendicular to each other \citep{Kennedy2019}. Numerical simulations suggest that this 'exotic' (yet stable) configuration can be reached in some multiple systems, the so-called polar configuration \citep{Verrier2008,Farago2010,Aly2018}. Dynamical evolution studies show that an inclined circumbinary disk around a highly eccentric (e $\gtrsim$ 0.7 ) inner binary can evolve towards this configuration \citep{Aly2015,Zanazzi2017,Cuello2019}. 

Recently, the orbital characterisation of hierarchical systems hosting disks has provided new insights on the mechanism involved in the formation of multiple systems and their interaction with the disk \citep{Kraus2020,Czekala2021}. To better understand the source of disk misalignment and the formation process behind hierarchical systems, better  information on well-characterised multiple systems' architectures will be necessary. In that regard, the full characterisation of the HD\,98800 quadruple system presents an opportunity to expand the sample of hierarchical systems hosting a protoplanetary disk.  

In this work, we present new long-baseline infrared interferometric observations of both AaAb and BaBb subsystems, as well as new RV measurements from original observations and archival reduced spectra. The new interferometric observations resolve the relative position of Ab with respect to Aa for the first time, providing one of the missing keys for the full characterisation of this quadruple system. Additionally, we also present two new astrometric positions for BaBb, allowing us to refine the orbital solution reported in \cite{Boden2005}. With the new orbital solutions of AaAb and BaBb, we re-estimated the orbital parameters of the AB outer orbit, evaluate the dynamical stability of this system, and discuss possible formation scenarios for this 2+2 quadruple.  

\section{Observations, astrometry, and RV}
    
\subsection{PIONIER observations and data reduction}
We used the Very Large Telescope Interferometer \citep[VLTI,][]{Haguenauer2008,Haubois2020} with the four-telescope combiner PIONIER in the H band \citep[$1.5\,-\,1.8\,\mu$m,][]{LeBouquin2011} to observe the HD\,98800 quadruple system. Our observations were carried out using the $1.8$\,m Auxiliary Telescopes with small and medium configurations, providing six projected baselines per configuration ranging from $\sim$ $20$ to $100$\,m. This configuration provides an angular resolution of $\sim4$ mas. The estimated interferometric field of view for PIONIER is $\sim160$ mas \citep{Hummel2016}, but given the loss of coherence caused by spectral smearing of the companion, with our given configuration, we have a field of view $\lesssim 60$\,mas \citep{Le-Bouquin2012,Gallenne2015}.

The first observations of both sub-systems were taken in April and May 2019 as a part of the science verification (SV) campaign\footnote{\tiny \url{https://www.eso.org/sci/activities/vltsv/naomisv.html}} of the New Adaptive Optics Module for Interferometry \citep[NAOMI,][]{Woillez2019}. These observations showcase the improvement provided by NAOMI on the sharpness of the point spread function (PSF) (despite $\sim 1$\arcsec~seeing conditions), which led to a better injection of the light in the fibre, and allowed us to mitigate light-contamination effects between A and B subsystems (A-B separation $ \lesssim \, 0.4$\arcsec). After the SV run, we obtained six more PIONIER epochs for the AaAb binary between February 2020 and March 2021 (see Table \ref{tab:astrometry}). 

\begin{figure*}
   \centering
   \includegraphics[width=0.99\textwidth]{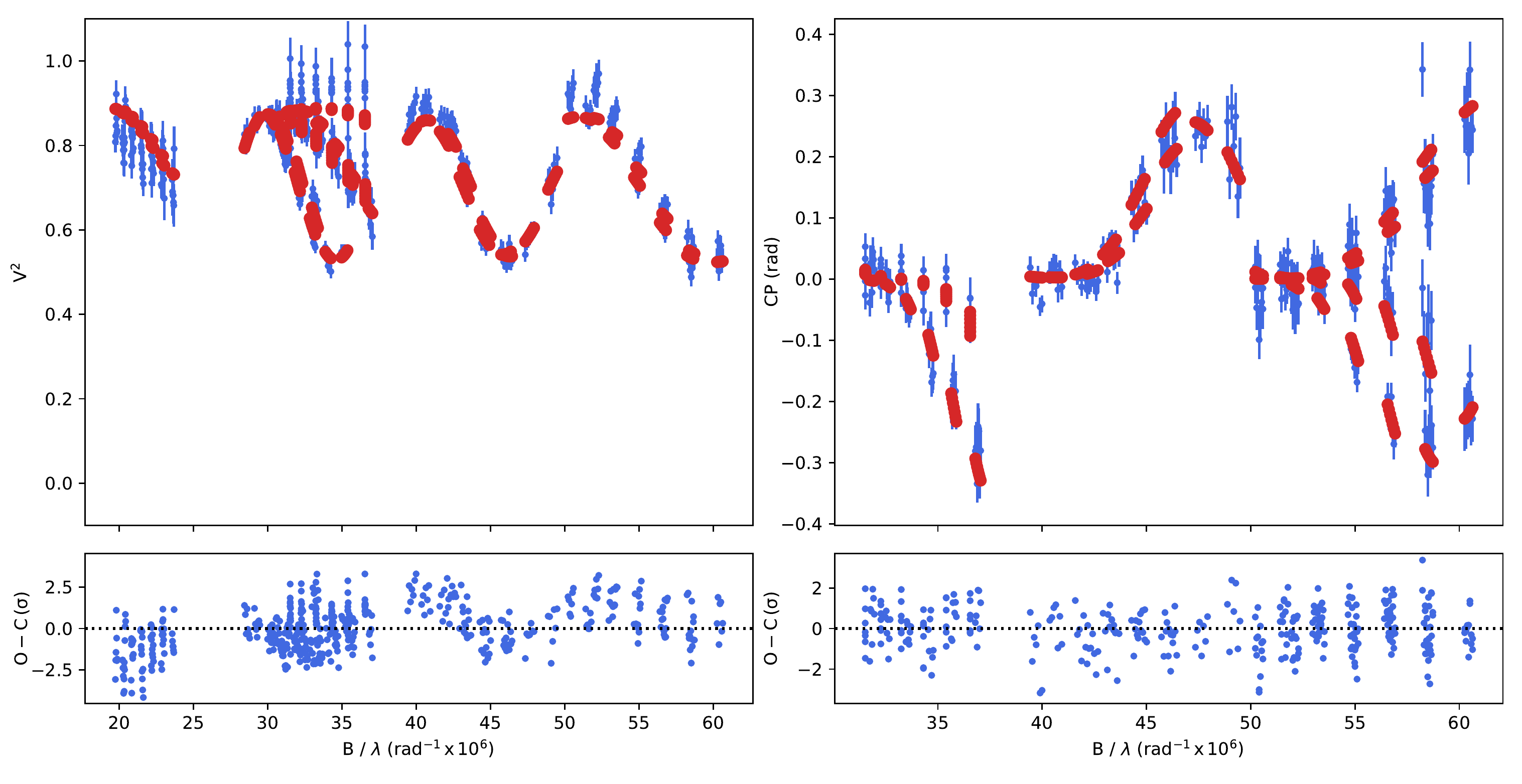}
   \caption{Squared visibility and closure phase measurements from one observation of AaAb taken in March 2021. The data are in blue, while the red dots represent the best binary model fitted with \texttt{CANDID} for this epoch. The bottom panels show the residuals in the number of sigmas.}\label{Fig:V2}%
    \end{figure*}

To monitor the instrumental and atmospheric transfer functions, the standard observing procedure is to interleave science and reference stars (CAL-SCI-CAL-SCI-CAL sequence). The calibrators, listed in Table \ref{tab:calibrators}, were selected using the \texttt{SearchCal} software \citep{Bonneau2006,Bonneau2011,Chelli2016} provided by the Jean-Marie Mariotti Center (JMMC\footnote{\tiny\url{https://www.jmmc.fr}}). The data were reduced with the \texttt{pndrs} package described in \cite{LeBouquin2011}. The main procedure is to compute squared visibilities ($V^2$) and triple products for each baseline and spectral channel, and to correct for photon and readout noise. The calibrated data are available in the Optical Interferometry DataBase\footnote{\tiny\url{http://oidb.jmmc.fr/index.html}}. In Fig. \ref{Fig:V2}, an example of the squared visibilities and closure phases (CP) for one of our observations of AaAb is presented.

    \begin{figure}
   \centering
   \includegraphics[width=0.49\textwidth]{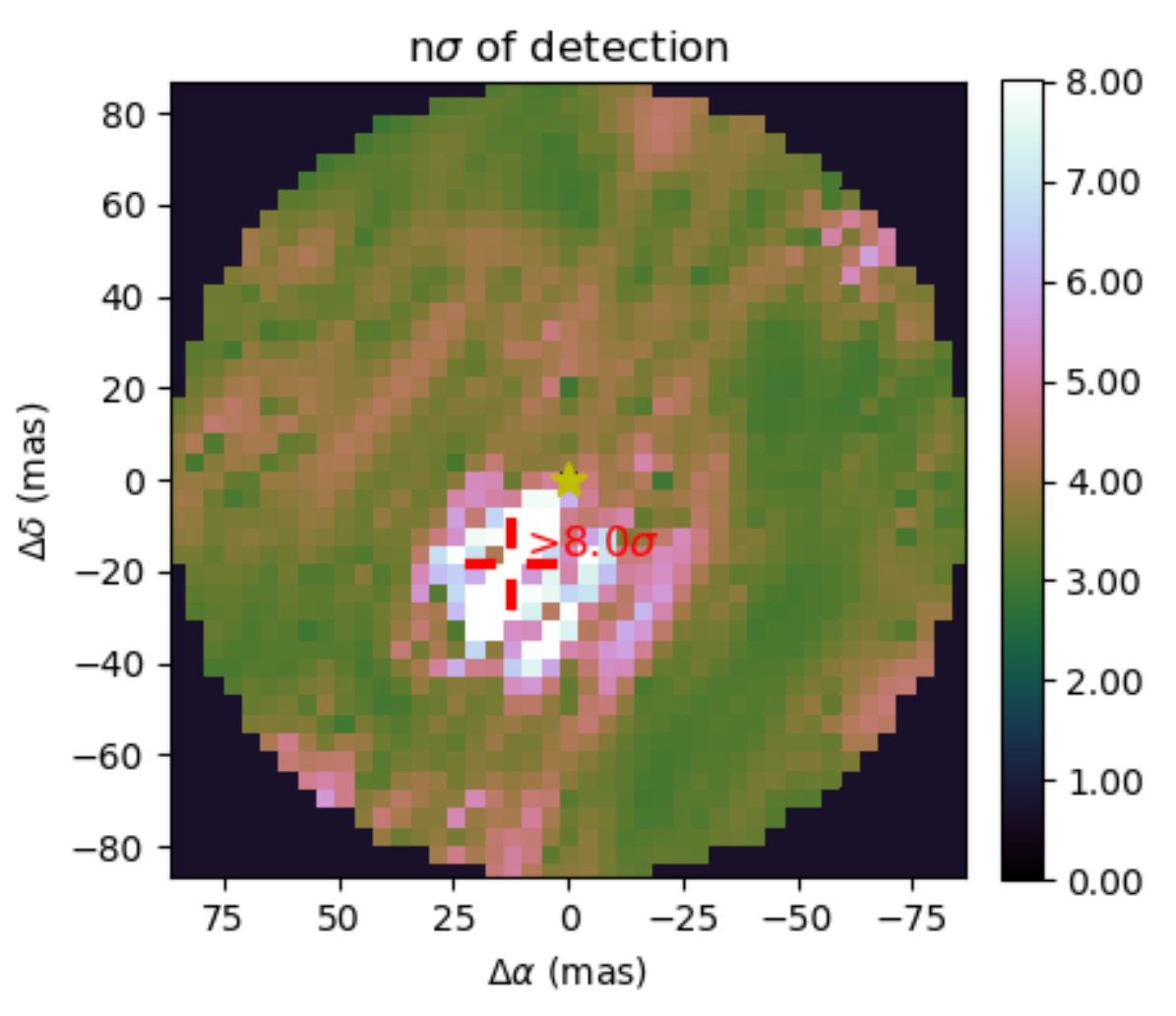}
   \caption{Detection level map from \texttt{CANDID} for the observation of AaAb taken in April 2019. The colourbar shows the significance of the companion detection in the number of sigmas.  The red cross points to the best-fit position.}\label{Fig:chi2Map}%
    \end{figure}

\subsection{Determining the AaAb and BaBb astrometry}

For each PIONIER observation, we determined the astrometric positions by fitting the  $V^2$ and CP with a binary model using the interferometric tool \texttt{CANDID}\footnote{\tiny\url{https://github.com/agallenne/GUIcandid}} \citep{Gallenne2015}. For each epoch, the tool delivered the binary parameters, namely the flux ratio ($f_2/f_1$) and the relative astrometric position ($\Delta\alpha$, $\Delta\delta$).  \texttt{CANDID} can also fit the angular diameter of both components, however, in our case, we kept them fixed at 0.3\,mas during the fitting process as the VLTI baselines did not allow us to resolve such small diameters. Briefly, the tool provides a 2D-grid of a multi-parameter fit using a least-squares algorithm (see Fig. \ref{Fig:chi2Map}). Given the small separation between AaAb and BaBb ($ \lesssim \, 0.4$ \arcsec), we also fitted an additional parameter to take the background cross-contamination into account, the non-coherent light, parametrised in \texttt{CANDID}  as a resolved flux ($f_{res}$). The final astrometric positions for all epochs of each subsystem are listed in Table \ref{tab:astrometry}. \texttt{CANDID} estimates the uncertainties using a bootstrapping approach (with replacement) using $10\,000$ bootstrap samples. For the flux ratio and resolved flux, we used the bootstrap sample distributions and took the median value as the best-fit result and the maximum value between the 16th and 84th percentiles as uncertainty. For the astrometry, the $1\sigma$ error region of each position is defined with an error ellipse parametrised with the semi-major axis  $\sigma_{maj}$ , the semi-minor axis $\sigma_{min}$ , and the position angle $\sigma_{PA}$ measured from north to east. We also quadratically added the systematic uncertainty of $0.35$\,\% from the precision of the PIONIER wavelength calibration to $\sigma_{maj}$ and $\sigma_{min}$
 \citep{Kervella2017,Gallenne2018}.
        
\begin{table*}[]
\tiny
                \centering
                \caption{Relative astrometric position of the secondary component, flux ratio, and resolved flux from PIONIER observations. The last two columns correspond to the atmospheric conditions for each epoch: the seeing and coherence time ($\tau_0$),  measured by the seeing monitor.}
                \begin{tabular}{crrrccccccc} 
                        \hline\hline \\
                        MJD &  $\Delta \alpha$  &  $\Delta \delta$      & $\sigma_\mathrm{PA}$    & $\sigma_\mathrm{maj}$ &  $\sigma_\mathrm{min}$        &  $f_2 / f_1$    & $f_{res}$\tablefootmark{a} & Baselines        & Seeing & $\tau_0$\\
                        &  (mas)                        & (mas)                 & ($^\circ$)      & (mas)         &  (mas)  &  (\%)  & (\%)&      & (arcsec) & (ms) \\
                        \multicolumn{11}{c}{}\\ 
                          \hline\\
                          \multicolumn{11}{c}{AaAb}  \\     
                          \hline\\  
                        58601.100162 & $12.38$ & $-18.35$ & $-13.68$ & $0.02$ &  $0.01$ &  $15.2\pm0.2$  & $8.6 \pm 0.5$  &  D0-G2-J3-K0 & $1.05$ & $4.98$\\ 
                        58615.047718 & 15.21 & -16.68 & $-75.72$ & $0.07$ &  $0.01$ &  $15.3\pm0.2$ &  $8.9\pm0.5$   &  A0-B2-C1-D0 & $1.08$ & $5.50$\\ 
                        58882.282610 & 15.80 & -16.10 & $68.64$ & $0.04$ &  $0.02$ &  $15.4 \pm 0.1$  &  $9.1\pm0.9$   &  A0-B2-C1-D0 & $0.93$ & $4.05$ \\ 
                        58899.329997 & 18.34 & -13.19 & $53.01$ & $0.02$ &  $0.01$ &  $14.6\pm0.3$  &  $ 7.2\pm1.1$   &  D0-G2-J3-K0 & $0.73$ & $6.23$ \\ 
                        58931.293872 & 20.32 & -5.61 & $-6.01$ & $0.03$ & $0.02$ &  $13.8\pm0.2$  &  $6.7\pm0.3$  &  D0-G2-J3-K0 & $0.58$& $10.01$ \\ 
                        59282.347290 & 6.26 & 14.37 & $-4.01$ & $0.01$ & $0.01$ &  $14.5\pm0.2$  & $9.3\pm0.7$   &  D0-G2-J3-K0 & $0.92$ & $6.19$ \\ 
                        59292.225719 & 2.72 & 14.33 & $21.17$ & $0.01$ & $0.01$ &  $ 15.3\pm0.3$  &  $8.7\pm0.6$  &  D0-G2-J3-K0 & $0.45$ & $5.14$\\ 
                        59295.208560 & 1.50 & 14.08 & $66.67$ & $0.01$ & $0.01$ &  $13.8\pm0.3$  &  $8.8\pm0.7$   &  D0-G2-J3-K0 & $0.81$ & $5.13$\\ 
                        \hline\\
                          \multicolumn{11}{c}{BaBb}  \\     
                          \hline\\
                        58601.106553 & 16.90 & -1.44 & -50.59 & 0.01 & 0.01  &  $70\pm1$  &  $12 \pm 1$  &  D0-G2-J3-K0  &$1.05$ & $4.98$ \\  
                        58615.013389 & 17.62 & -3.40 & 33.15 & 0.03 & 0.01  &  $65 \pm 1$  &  $13\pm1$  &  A0-B2-C1-D0 & $1.08$ & $5.50$ \\ 
                        \hline
                \end{tabular}
                \tablefoot{\tablefoottext{a}{Parameter to take the background cross-contamination into account (non-coherent light), parametrised in \texttt{CANDID}  as a resolved flux.}}
                \label{tab:astrometry}
        \end{table*}

\subsection{AB astrometry}

We gathered astrometric measurements from the Washington Double Star catalogue \citep[WDS,][]{WDS2001}. The AB pair has been observed since 1909, and observations before 1991 have no reported uncertainties. For those observations, the expected astrometric uncertainty was found to be between $0.02\,-\,0.1$\,\arcsec, depending on the target brightness and observing conditions \citep{Douglass1992,Torres1999}. Since 2009, the pair has been regularly observed with the speckle camera \citep[HRCam,][]{SI_Tokovinin2018} mounted on the $4.1$\,m Southern Astrophysical Research Telescope (SOAR); the last observation presented in this work was obtained in April 2021.

\subsection{CTIO spectroscopy}

Five observations were taken with the $1.5$\,m telescope located at the Cerro Tololo Inter-American Observatory (CTIO) in Chile, and operated by the Small and Moderate Aperture Research Telescopes System (SMARTS) Consortium\footnote{\tiny\url{ http://www.astro.yale.edu/smarts/}}, from April-July 2021. Observations were made with the CHIRON optical echelle spectrograph \citep{Tokovinin2013}. The RVs were determined from the cross-correlation function (CCF) of echelle orders with the binary mask based on the solar spectrum, as detailed in \cite{Tokovinin2016a}. From these observations, we obtained five RV measurements for Aa (brighter component). The Ba and Bb components were totally blended with Aa at two epochs and they could not be separated by a multi-component fit. However, the blending certainly biases the RVs of Aa, increasing the uncertainty of these measurements. In three observations, we were able to obtain reliable RV measurements for Ba and Bb; in one of them, however, the components were still partially blended, so a larger uncertainty was assigned to it. 

\subsection{Reduced spectra from public archives}

We found nine science-ready datasets in the ESO Phase 3 public archive\footnote{\tiny\url{http://archive.eso.org/scienceportal/home}} taken with the Fibre-fed Extended Range \'Echelle Spectrograph (FEROS/2.2\,m, \citealt{FEROS1999}). The 1D Phase 3 spectra are given in the barycentric reference frame. One observation was taken in 2015 and the remaining eight were acquired between July and August 2007. The RVs were determined by cross-correlation with the same solar-type binary mask as used in CHIRON. The lines are blended and dominated by the lines of Aa. Consequently, with these FEROS spectra, we obtained only RV measurements for the Aa component, potentially biased by blending with Ba and Bb. Additionally, we found two reduced spectra in the ELODIE public archive\footnote{\tiny\url{http://atlas.obs-hp.fr/elodie/}} at the Observatoire de Haute-Provence \citep[OHP,][]{ELODIE_Archive2004}. The observations were taken in 1998, on January 28 and 29. The spectra are not given in the barycentre reference frame; a correction was therefore applied after retrieving the data. The RVs were determined from the CCF of the spectra with a CORAVEL-type G2 numerical mask using a standalone CCF tool \footnote{\tiny\url{https://github.com/szunigaf/CCF_functions}} \citep[for further details, see][]{Zuniga-Fernandez2021}. The spectra from ELODIE show partially blended lines and were fitted by three Gaussian components. The RV measurements for BaBb from ELODIE have large uncertainties, but still allowed us to compute the systemic velocity of BaBb at this epoch.

\subsection{Literature data}
From the literature, we collected a diverse dataset for this system. The RV measurements of the primary star of AaAb (single-line spectroscopic system, SB1) and both components for BaBb (double-line spectroscopic system, SB2) were taken from \cite{Torres1995}, hereafter TO95. For the BaBb binary, we also retrieved interferometric $V^2$ measurements, obtained with the KI, and published in \cite{Boden2005}. Additionally, assuming that the RV of B is the same as the systemic velocity of the disk, we include the disk RV derived from the CO modelling presented in \cite{Kennedy2019}, hereafter KE19.

\section{Orbital fitting}
\begin{figure}
   \centering
   \includegraphics[width=0.49\textwidth]{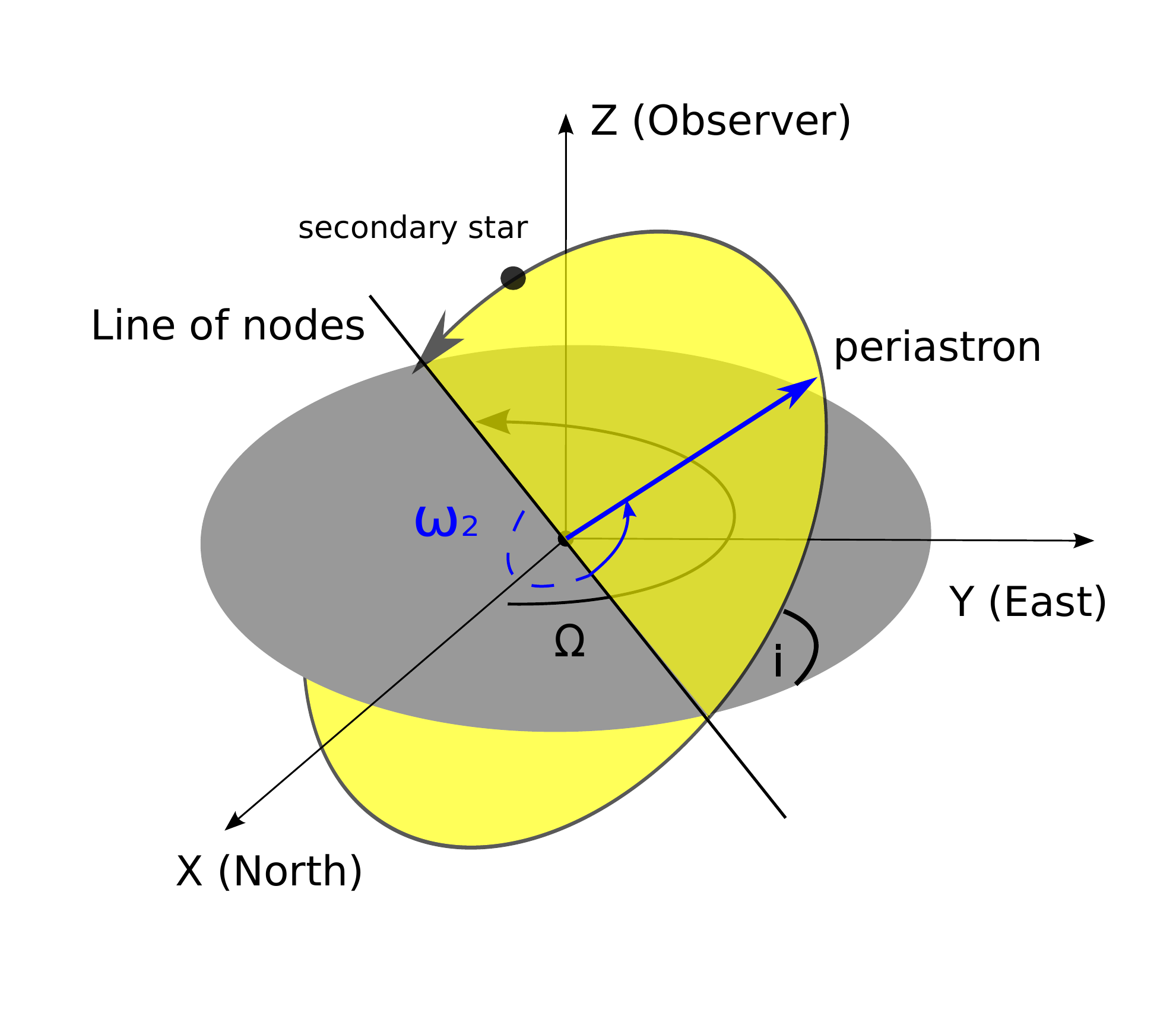}
   \caption{Diagram of the orbit of the secondary star around the centre of mass (yellow plane) and the reference plane (grey). This diagram follows the orbital convention of \texttt{exoplanet}.}
   \label{Fig:orbit_conv}
    \end{figure}

We modelled our dataset with the \texttt{exoplanet} software package \citep{exoplanet:exoplanet}, which extends the \texttt{PyMC3} framework \citep{exoplanet:pymc3} to support many of the custom functions and distributions required when fitting orbital parameters. Some of the parameters describing the primary or the secondary star orbits around the centre of mass are identical for both components, for example, the period (P), eccentricity (e), inclination (i), and longitude of the ascending node ($\Omega$). But others depend on the component used as a reference, for example, the semi-amplitude of the RV ($K_{~\mathrm{primary}}$ and $K_{~\mathrm{secondary}}$) and the argument of the periastron ($\omega_{~\mathrm{primary}}\,=\,\omega_{~\mathrm{secondary}}+ \pi$). In an astrometric-only orbital fitting, it is common practice to report $\omega\,=\,\omega_{~\mathrm{secondary}}$, whereas with an RV orbit it is generally common practice to report $\omega\,=\,\omega_{~\mathrm{primary}}$. Then, in a joint astrometric-RV orbit, there could be ambiguity regarding the convention used for $\omega$. We adopted the orbital convention from  \texttt{exoplanet}\footnote{\tiny\url{https://docs.exoplanet.codes/en/latest}}, where the argument of periastron $\omega$ is reported with respect to the primary star, and the longitude of the ascending node $\Omega$ is the node where the secondary is moving away from the observer (see Fig. \ref{Fig:orbit_conv}). 

Given that BaBb is an SB2, the orbital fitting procedure is slightly different compared to the AaAb subsystem (SB1). The different steps for each orbital fitting are explained below. The Markov chain Monte Carlo (MCMC) samples and PyMC3 models corresponding to both subsystems are available online\footnote{\tiny\url{https://github.com/szunigaf/HD98800}}. The prior distributions and corner plots from the orbital parameters' posterior samples are displayed in Appendix \ref{sec:apendix_orbits}. 

\subsection{BaBb orbit}

Given that BaBb is an SB2, we can fit the astrometric points from PIONIER together with the RV amplitude of each component, $K_{\mathrm{Ba}}$ for the primary and $K_{\mathrm{Bb}}$ for the secondary, as well as the systemic velocity $\gamma_B$. Additionally, we extended \texttt{exoplanet} to include the $V^2$ model for individually unresolved components in a binary system. Briefly, the fringe contrast $V^2$ of a binary system depends on the properties of the individual components and the binary separation \citep{Berger2007},

\begin{equation}
    V_{~\mathrm{binary}}^2 = \frac{ 1 + \left(\frac{f_2}{f_1}\right)^2 + 2 \left(\frac{f_2}{f_1}\right) \, \cos{ \left( \frac{2\, \pi \,C \,(u \,\Delta \alpha  + v \, \Delta \delta)} {\lambda} \right) }} { \left(1 + \frac{f_2}{f_1}\right)^2},
\end{equation}

\noindent where $\Delta \alpha$  and $\Delta \delta$ are the relative separation in right ascension and declination, respectively (from the \texttt{exoplanet} model), $u$ and $v$ are the projected baselines (in meters), ${f_2}/{f_1}$ is the flux ratio, and $\lambda$ is the wavelength. The parameter $C$ is a conversion factor so that the astrometry is in arcsec and the wavelength in $\mu$m. The $V^2$ measurements from KI were then included in the fitting process, where the flux ratio ${f_2}/{f_1}$ was fitted as a free parameter along with the other orbital parameters (see Table \ref{tab:orbit_BaBb}).

\begin{figure}
   \centering
   \includegraphics[width=0.48\textwidth]{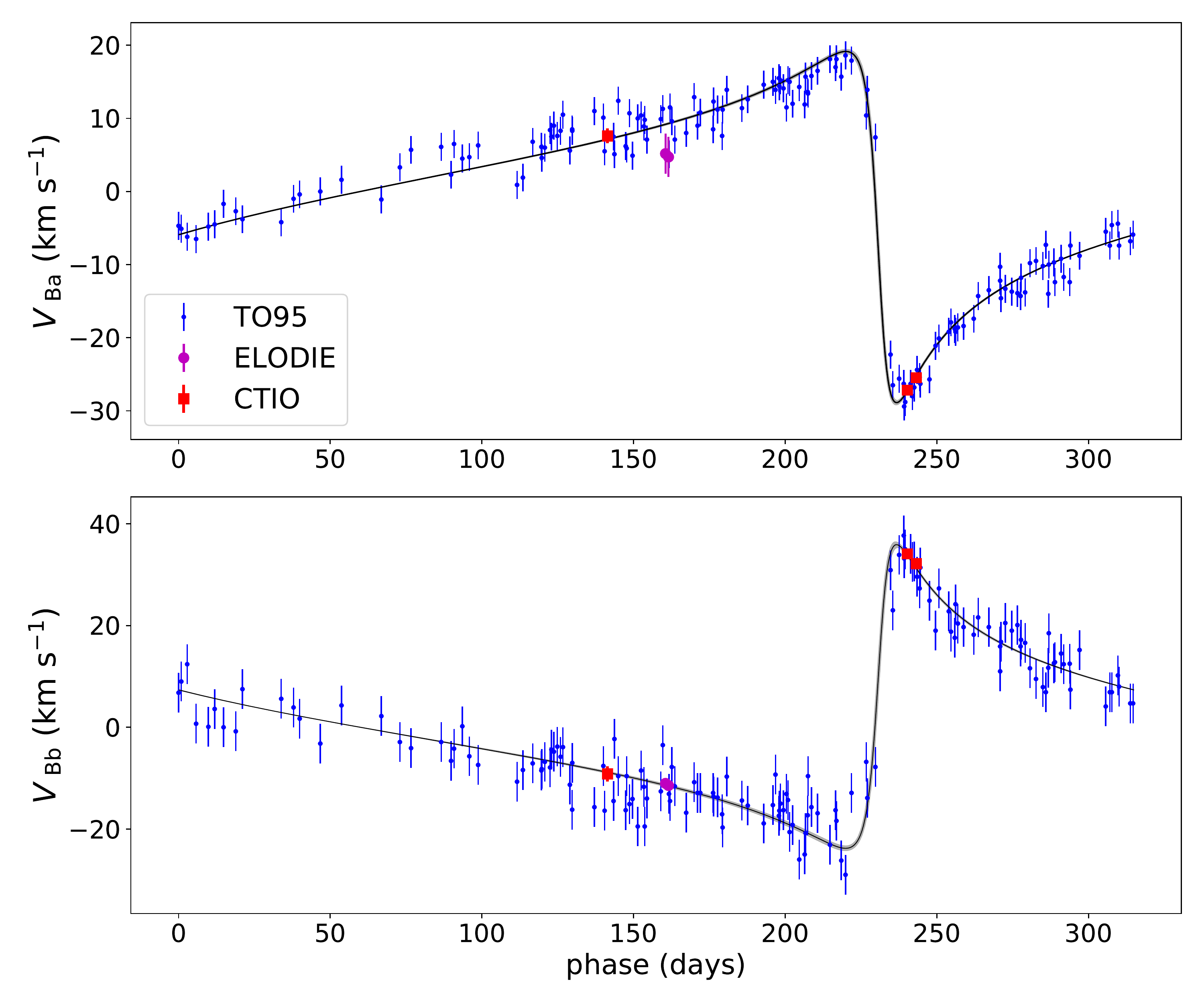}
   \caption{Phase-folded RVs orbit for BaBb. The systemic velocity $\gamma$ for each set of observations was subtracted. The solid line corresponds to the best-fit model. The upper panel plots the RVs of Ba, and the lower panel corresponds to Bb.} \label{Fig:BaBa_RVorbit}%
    \end{figure}

\begin{figure}
   \centering
   \includegraphics[width=0.48\textwidth]{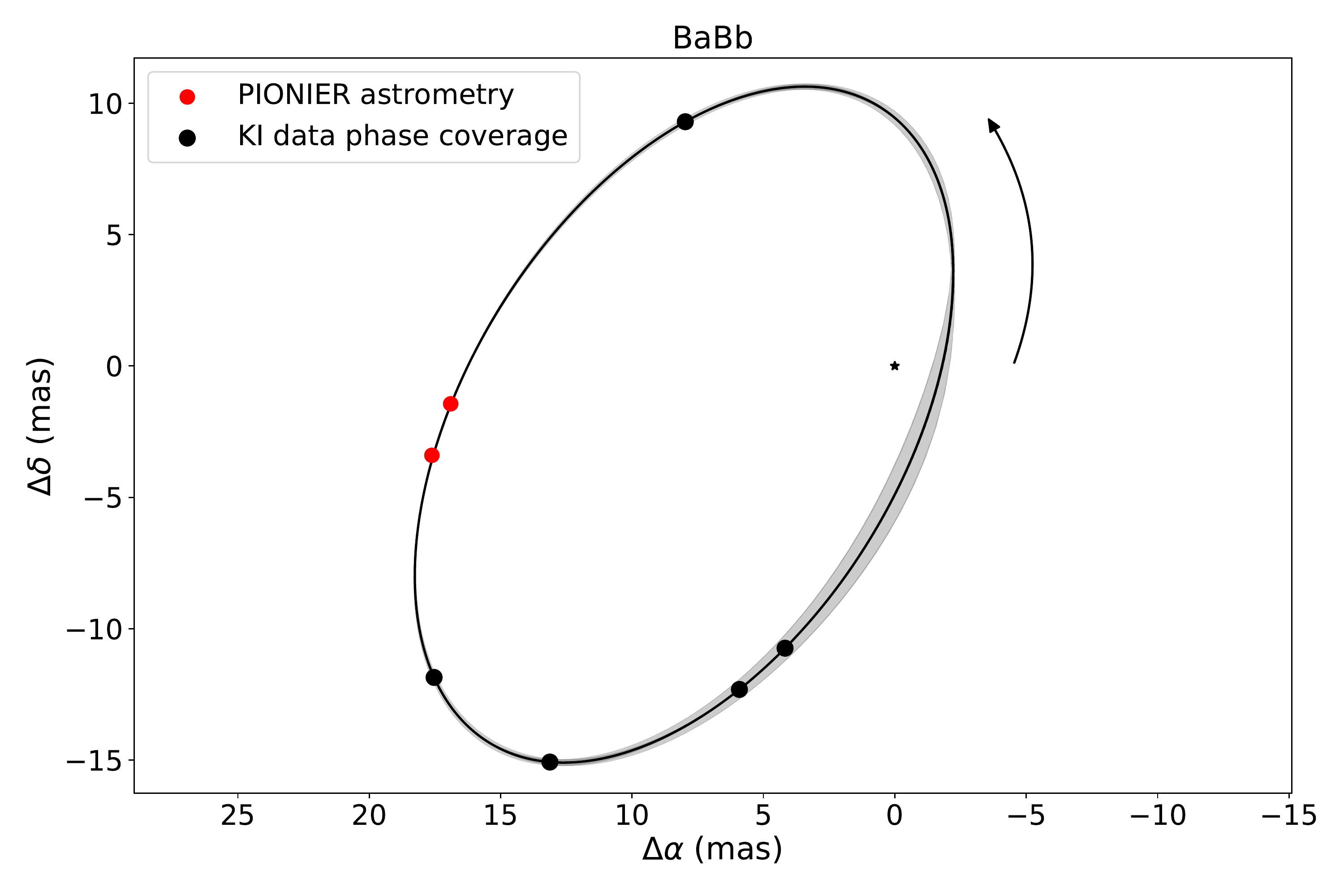}
   \caption{Best orbital solution for BaBb. The solid line corresponds to the best-fit model and the shaded area to the $1\sigma$ region. The primary star Ba is located at the origin. The relative positions of Bb are plotted as filled dots. The error ellipses from PIONIER astrometry are smaller than the markers.}\label{Fig:BaBa_orbit}%
    \end{figure}

All the orbital parameters were estimated from the posterior distributions, taking the median values as the best-fit results and the maximum values between the 16th and 84th percentile as uncertainties. From these distributions, we could then calculate the distribution of the masses for both components as well as the distance to the system using the following equations \citep{Torres2010,Gallenne2019}:

\begin{eqnarray}
                M_{Ba} &=& \dfrac{1.036149\times 10^{-7} (K_{Ba} + K_{Bb})^2 K_{Bb}\,P\,(1 - e^2)^{3/2} }{\sin ^3 i}, \label{eq:M1}\\
                M_{Bb} &=& \dfrac{1.036149\times 10^{-7} (K_{Ba} + K_{Bb})^2 K_{Ba}\,P\,(1 - e^2)^{3/2} }{\sin ^3 i}, \label{eq:M2} \\
                a_\mathrm{au} &=& \dfrac{9.191940\times 10^{-5} (K_{Ba} + K_{Bb})\,P \sqrt{1 - e^2} }{\sin i}, \label{eq:a_au}\\
                \pi &=& \dfrac{a}{a_\mathrm{AU}},
                \label{eq:plx}
\end{eqnarray}

\noindent where $M_{Ba}$ and $M_{Bb}$ correspond to the masses of the primary and the secondary stars, respectively, expressed in solar mass, $P$ is the period in days, $K_{Ba}$ and $K_{Bb}$ are the RV semi-amplitudes of the primary and secondary star in \kms, respectively, and $a$ is the angular semi-major axis in arcseconds. The parameter $a_\mathrm{au}$ is the semi-major axis expressed in astronomical units. Table \ref{tab:orbit_BaBb} lists a full description of the inferred orbital parameters. Fig. \ref{Fig:BaBa_RVorbit} shows the best-fit RV curve. Fig. \ref{Fig:BaBa_orbit} shows the best-fit visual orbit; the black dots are the phase coverage of the KI observations, that is the astrometric positions from the best-fit orbit corresponding to the observation date of each $V^2$ dataset  (see Fig. \ref{fig:KI_model}). Some parameters seem incompatible with the previous result taking the uncertainties into account; this may be due to the fact that some of the uncertainties could have been underestimated.

\begin{table}[] 
        \small
                \centering
                \caption{Orbital parameters for the HD 98800 BaBb binary.}
                \begin{tabular}{lrr}
                        \hline
                        \hline  \\
                        Orbital parameters &   \cite{Boden2005} & This work \\ 
                          \hline\\
                        Period (days)      & 314.327  $\pm$ 0.028  & 314.86 $\pm$ 0.02 \\ 
            T$_{0}$ (MJD)      & 52481.34 $\pm$ 0.22   & 48707.5 $\pm$ 0.2 \\ 
            $e$                   & 0.7849 $\pm$ 0.0053   & 0.805$\pm$0.005 \\ 
            $\omega_{Ba}$ ($^\circ$)     & 109.6 $\pm$ 1.1       & 104.5 $\pm$ 0.3 \\ 
            $\Omega$ ($^\circ$)          & 337.6 $\pm$ 2.4       & 342.7 $\pm$ 0.4 \\ 
            $i$ ($^\circ$)                  & 66.8  $\pm$ 3.2       & 66.3 $\pm$ 0.5 \\
            $a$ (mas)                 & 23.3  $\pm$ 2.5       & 22.2 $\pm$ 0.4 \\
            K$_{Ba}$ (km s$^{-1}$)  & 22.94 $\pm$ 0.34      &  24.0 $\pm$ 0.3 \\ 
            K$_{Bb}$ (km s$^{-1}$)    & 27.53 $\pm$ 0.61      & 29.9 $\pm$ 0.6 \\ 
            $\gamma_{~\mathrm{TO95}}$ (km s$^{-1}$)  & 5.73  $\pm$ 0.14 & 5.6 $\pm$ 0.1 \\ 
            $\gamma_{~\mathrm{ELODIE}}$ (km s$^{-1}$)  & \dots & 3.4 $\pm$ 0.7 \\ 
            $\gamma_{~\mathrm{CTIO}}$ (km s$^{-1}$)  & \dots & 6.4 $\pm$ 0.4 \\ 
            $f_2/f_1$   (K band)               &  0.612 $\pm$ 0.046    &  0.76 $\pm$ 0.08  \\ 
            & &  \\
            \hline
            \hline \\
            Derived parameters & &  \\
            \hline\\ 
            $\pi$ (mas)              & 23.7 $\pm$ 2.6     &  22.0 $\pm$ 0.6 \\
            $M_{Ba}$ (M$_{\sun}$)    & 0.70 $\pm$ 0.06    &  0.77 $\pm$ 0.04 \\
            $M_{Bb}$ (M$_{\sun}$)     & 0.58 $\pm$ 0.05     & 0.62 $\pm$ 0.02 \\
            d (pc) & 42.2 $\pm$ 4.7 & 45 $\pm$ 1 \\
            a (AU) & 1.0 $\pm$ 0.2 & 1.01 $\pm$ 0.01 \\
                        \hline
                \end{tabular}
                \label{tab:orbit_BaBb}
        \end{table} 

\subsection{AaAb orbit}
\label{sec:AaAb-Orbit}
This subsystem is an SB1, therefore it is not possible to break the degeneracy between the parallax and the semi-major axis and determine individual stellar masses. The orbit is based on the astrometric points from PIONIER  and on the RVs of the primary star Aa. Consequently, we only fitted the RV semi-amplitude of the Aa component of the system $~\mathrm{K_{Aa}}$, and the systemic velocity $\gamma_A$. To estimate the masses of the individual components of an SB1, we must assume a distance. We tested two parallax values, the one obtained from the orbital fitting of BaBb and the Hipparcos one \citep{Hipparcos2007}, as there is currently no reliable Gaia parallax for the system. In our MCMC model, we included these parallax values as a prior using a normal distribution ($22.27\pm2.31$\,mas and $22.0\pm0.6$ for the Hipparcos one and the one based from the orbital solution of BaBb, respectively). The parallax is then a free parameter in our orbital fitting using the abovementioned priors.  Using Kepler's third law and combining equations (\ref{eq:M2}) and (\ref{eq:a_au}), we calculated
\begin{eqnarray}
        M_{tot} &=& \dfrac{a_{AU} ^3}{P_{\mathrm{years}} ^2}, \label{eq:Mtot} \\
        M_{Ab} &=& \dfrac{1.036149\times 10^{-7}\,K_{Aa}\,\sqrt{1 - e^2}\,a_{AU} ^2}{(9.191940\times 10^{-5})^2\,P\,\sin i}, \label{eq:MAb} \\
        M_{Aa} &=& M_{tot} - M_{Ab}, \label{eq:MAa}
\end{eqnarray}

\noindent where $M_{tot}$, $M_{Aa}$,  and $M_{Ab}$ correspond to the total mass, and the primary and secondary star masses, respectively, expressed in solar mass, $P_{\mathrm{years}}$ the period in years, $P$ the period in days, $K_{Aa}$ the RV semi-amplitude of the primary star in \kms, and $a_{AU}$ the semi-major axis expressed in astronomical units.

\begin{figure}[]
   \centering
   \includegraphics[width=0.48\textwidth]{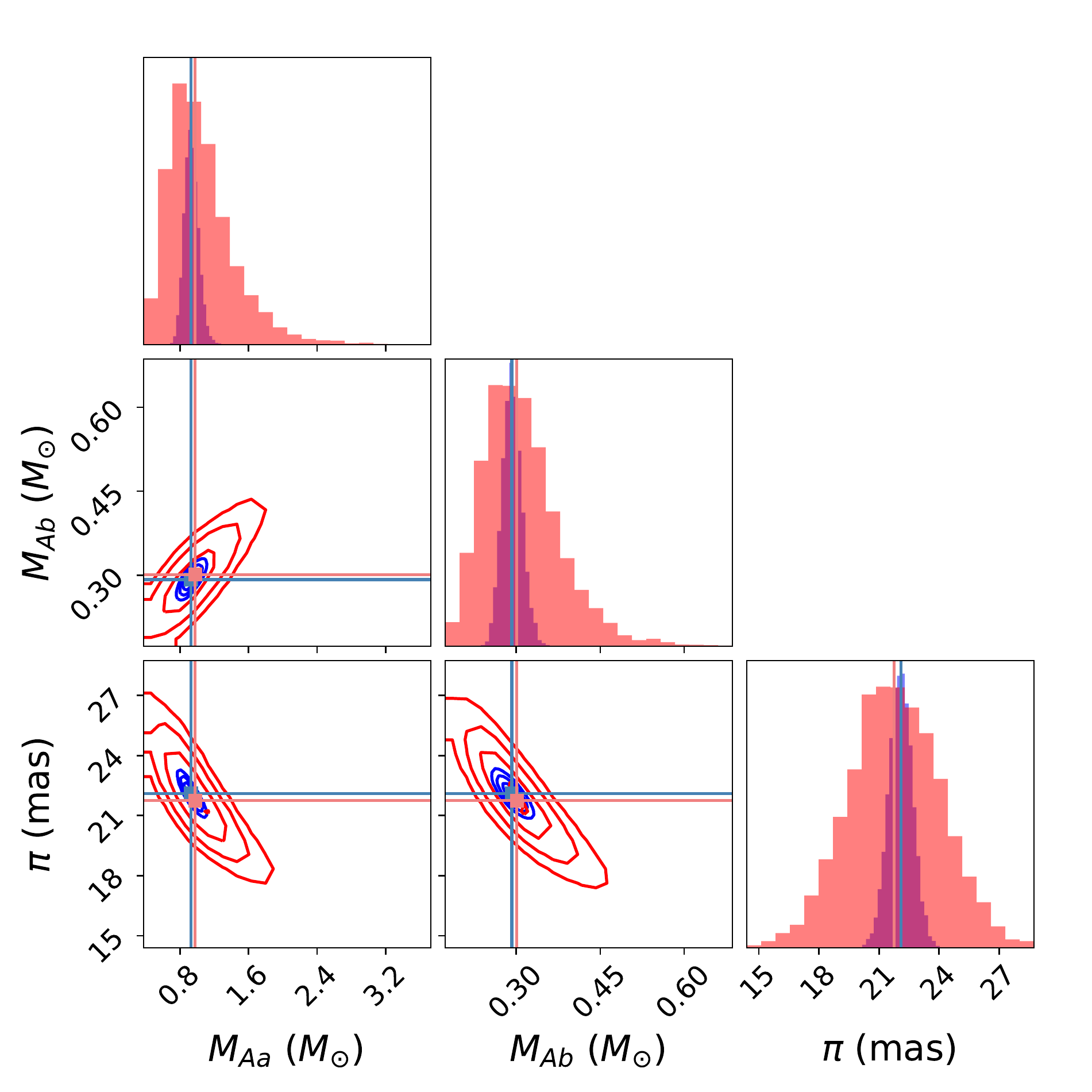}
   \vspace{-0.2cm}
   \caption{Posterior distribution of masses and parallax of AaAb subsystem assuming the Hipparcos value (red) and the solution from BaBb fitting (blue) as a prior distribution of the parallax in our MCMC model. The red and blue lines highlight the median of each distribution.}
  \label{Fig:AaAb_mass}%
    \end{figure}

The posterior distributions for the masses and parallaxes, assuming different initial priors for the parallaxes are shown in Fig. \ref{Fig:AaAb_mass} in blue and red, respectively. The orbital parameters converge and have the same results for both cases, only the physical parameters that are dependent on the distance are affected by the choice of the prior distribution for the parallax (i.e. $M_{Aa}$, $M_{Ab}$, and $a_{AU}$). 

\begin{figure}
   \centering
   \includegraphics[width=0.48\textwidth]{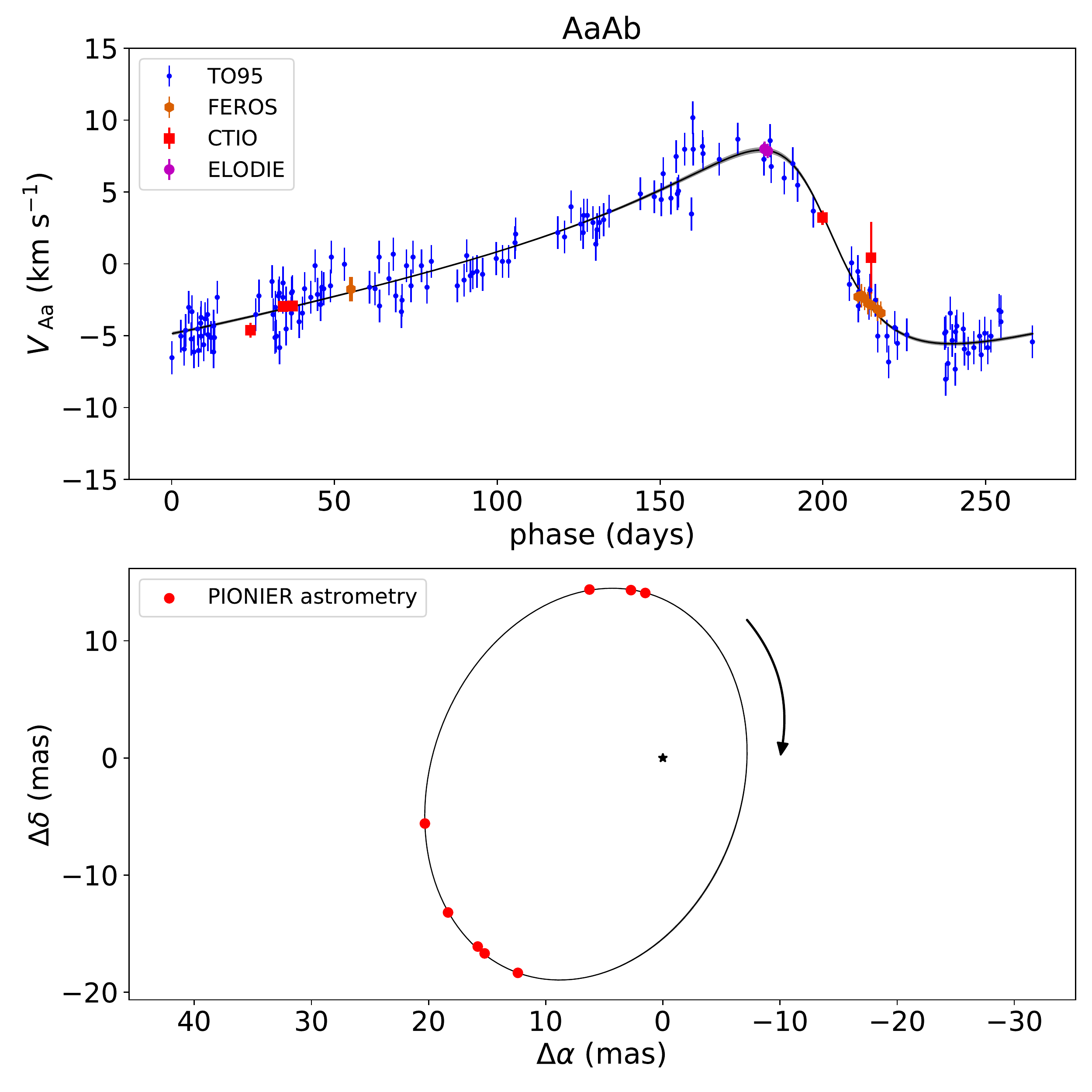}
   \vspace{-0.25cm}
   \caption{Best orbital solution for AaAb. In both panels, the solid line corresponds to the best-fit model. \textit{Bottom panel:} The primary star Aa is located at the origin. The relative positions of Ab are plotted as filled dots; the error ellipses from PIONIER astrometry are smaller than the marker. \textit{Upper panel:} The coloured markers correspond to the primary star RV measurements. The systematic velocity $\gamma$ for each set of observations was subtracted.} 
  \label{Fig:AaAb_orbit}%
    \end{figure}

All the orbital parameters were estimated from the posterior distributions taking the median values as the best-fit results and the  maximum  values  between  the  $16$th  and  $84$th  percentile  as uncertainties (see Table \ref{tab:orbit_AaAb}). Fig. \ref{Fig:AaAb_orbit} shows the best-fit binary orbit (identical in the plane of the sky for both parallax scenarios). In the rest of the paper, we assume the masses of AaAb as derived with the parallax obtained from the BaBb best orbital fit.

    \begin{table}[]
    \small
                \centering
                \caption{Orbital parameters for HD 98800 AaAb binary.}
                \begin{tabular}{lrr}
                        \hline
                        \hline  \\
                        Orbital parameters & \cite{Torres1995} &  This work \\
                          \hline\\
                        Period (days)               & 262.15 $\pm$ 0.51    & 264.51 $\pm$ 0.02  \\
            T$_{0}$ (MJD)            & 48737.1 $\pm$ 1.6   &  48742.5 $\pm$ 0.8 \\
            $e$                      & 0.484 $\pm$ 0.020    &  0.4808 $\pm$ 0.0008 \\
            $\omega_{Aa}$ ($^\circ$)      & 64.4 $\pm$ 2.1          &  68.7 $\pm$ 0.1  \\
            $\Omega$ ($^\circ$)           &      \dots                & 170.2 $\pm$ 0.1  \\
            $i$ ($^\circ$)                &  \dots         & 135.6 $\pm$ 0.1 \\
            $a$ (mas)                &  \dots                  &  19.03 $\pm$ 0.01 \\
            K$_{Aa}$ (km s$^{-1}$)   & 6.8 $\pm$ 0.2       &  6.7 $\pm$ 0.2 \\
            $\gamma_{~\mathrm{TO95}}$ (km s$^{-1}$)   & 12.7 $\pm$ 0.1    & 12.8 $\pm$ 0.1\\
            $\gamma_{~\mathrm{FEROS07}}$\tablefootmark{a}  (km s$^{-1}$)   & \dots    &  14.7 $\pm$ 0.4 \\
             $\gamma_{~\mathrm{FEROS15}}$\tablefootmark{b}  (km s$^{-1}$)   & \dots    &  12 $\pm$ 2 \\
            $\gamma_{~\mathrm{CTIO}}$ (km s$^{-1}$)   & \dots &  11.8 $\pm$ 0.2 \\
            $\gamma_{~\mathrm{ELODIE}}$ (km s$^{-1}$)   & \dots &  12.1 $\pm$ 0.5 \\
            & & \\
            \hline
            \hline \\
            Derived parameters & & \\
            \hline\\ 
            Hipp. $\pi$ (mas)            &    \dots       &  22 $\pm$ 2 \\
            $M_{Aa}$ (M$_{\sun}$)        &    \dots   &   0.9 $\pm$ 0.4 \\
            $M_{Ab}$ (M$_{\sun}$)        &    \dots    & 0.29 $\pm$ 0.07 \\
            a (AU) & \dots & 0.9 $\pm$ 0.1 \\
                        \hline\\ 
            BaBb $\pi$ (mas)            &    \dots    &  22.0 $\pm$ 0.6 \\
            $M_{Aa}$ (M$_{\sun}$)        &   \dots    &  0.93 $\pm$ 0.09 \\
            $M_{Ab}$ (M$_{\sun}$)        &   \dots    &  0.29 $\pm$ 0.02\\
            a (AU) & \dots & 0.86 $\pm$ 0.02 \\
                        \hline
                \end{tabular}
                \tablefoot{\tablefoottext{a}{Systemic velocity of FEROS observations taken in 2007.}\tablefoottext{b}{Same as (a), but for the FEROS observation taken in 2015.}}
                \label{tab:orbit_AaAb}
        \end{table}

\subsection{Outer orbit A-B}

Using our orbital solutions of the inner binaries of the system, we recalculated the orbital parameters of AB. We assume that the systemic velocities of AaAb and BaBb from \cite{Torres1995}, FEROS, CHIRON, and ELODIE observations in our orbital fitting results, and the one from CO modelling by \cite{Kennedy2019} (KE19), correspond the centre-of-mass  RVs of A and B in the outer orbit (see Table \ref{tab:appendix_RV_AB}). We jointly fitted the astrometric position with the RV measurements of AB. In our MCMC model, we included the parallax and the masses obtained from the inner orbits' results as  priors. For consistency, we used the AaAb masses derived from the parallax obtained from the orbital fitting of BaBb. The normal distribution priors for the masses and parallax are $M_A: 1.22\,\pm\,0.5\,\mathrm{M_{\sun}}$, $M_B: 1.38\,\pm\,0.5\,\mathrm{M_{\sun}}$, and $\pi: 22.0\,\pm\,0.6\,\mathrm{mas}$, respectively. The $\gamma_{~\mathrm{AB}}$ was included as a free parameter, with a uniform prior between 0 and 20 \kms. Given that the visual micrometric measurements made before 1991 have unknown uncertainties, we defined the large ($\sigma\sim0.1$\arcsec) and the small ($\sigma\sim0.02$\arcsec) uncertainty cases for these measurements (solutions I and II in Table \ref{tab:orbit_AB}), according to the typical range of errors reported in the astrometry measurements by USNO \citep{Douglass1992,Torres1999}.

\begin{figure}[]
   \centering
   \includegraphics[width=0.46\textwidth]{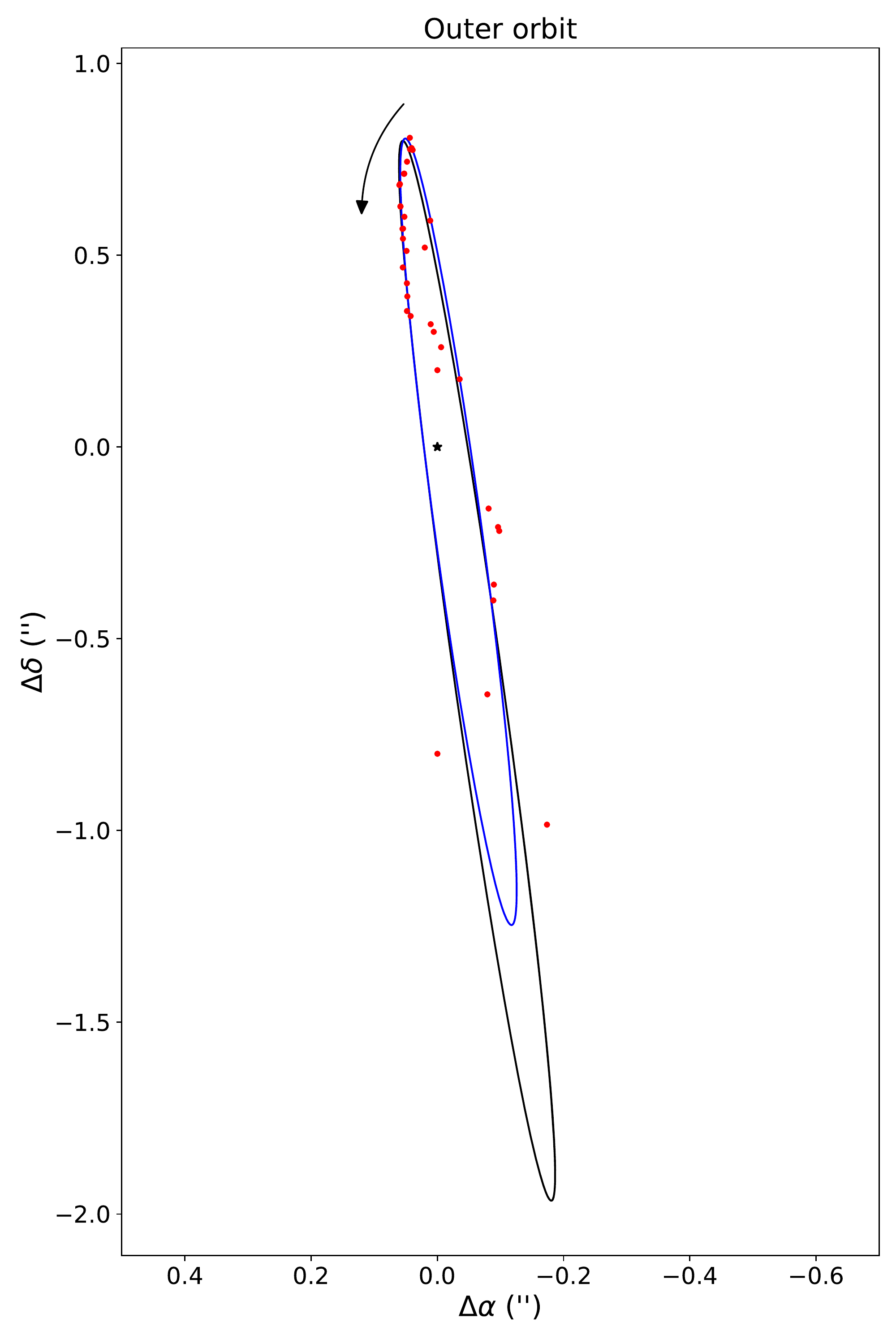}
   \vspace{-0.2cm}
   \caption{Best orbital solution for AB outer orbit for both uncertainty assumptions in the astrometry before 1991. The solid black line corresponds to the best orbital solution assuming small uncertainties ($\sigma\sim0.02$\arcsec) and the blue one assumes large uncertainties ($\sigma\sim0.1$\arcsec). The red dots correspond to the astrometric measurements. For better visualisation, the error bars are shown only in Fig. \ref{Fig:AB_orbit_sepPA}.}
  \label{Fig:AB_orbit}%
    \end{figure}

\begin{figure}[]
   \centering
   \includegraphics[width=0.46\textwidth]{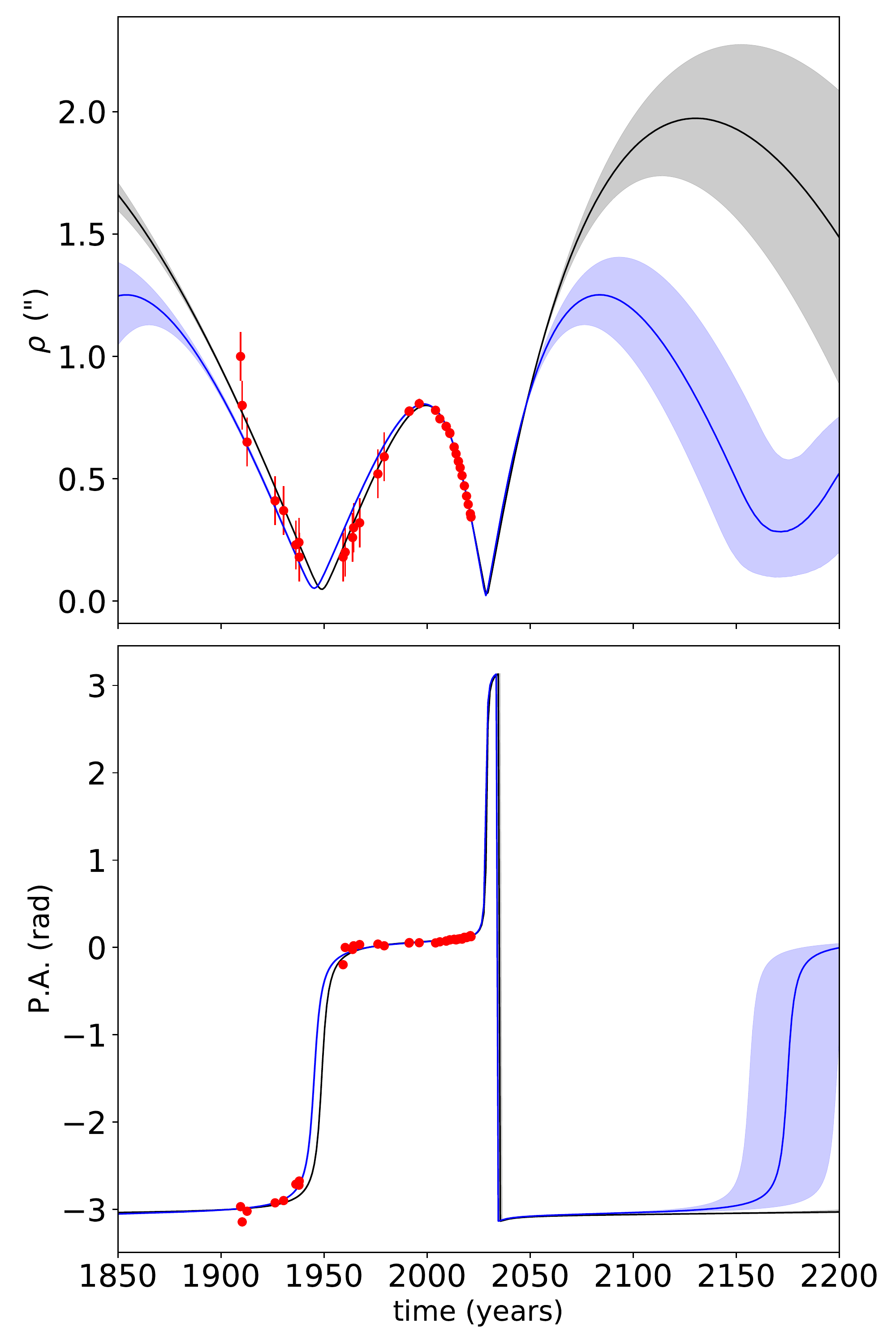}
   \vspace{-0.2cm}
   \caption{Best orbital solution for AB outer orbit for both uncertainty assumptions in the astrometry before 1991. In all panels the solid line corresponds to the best-fit model and the shaded area to the $1\sigma$ region. The solid black lines correspond to the best orbital solution assuming small uncertainties ($\sigma\sim0.02$\arcsec) and the blue ones assume large uncertainties ($\sigma\sim0.1$\arcsec). The red dots correspond to the astrometric measurements. The error bars shown in the astrometry before 1991 correspond to the large uncertainty case.}
  \label{Fig:AB_orbit_sepPA}%
    \end{figure}

\begin{figure}[]
   \centering
   \includegraphics[width=0.49\textwidth]{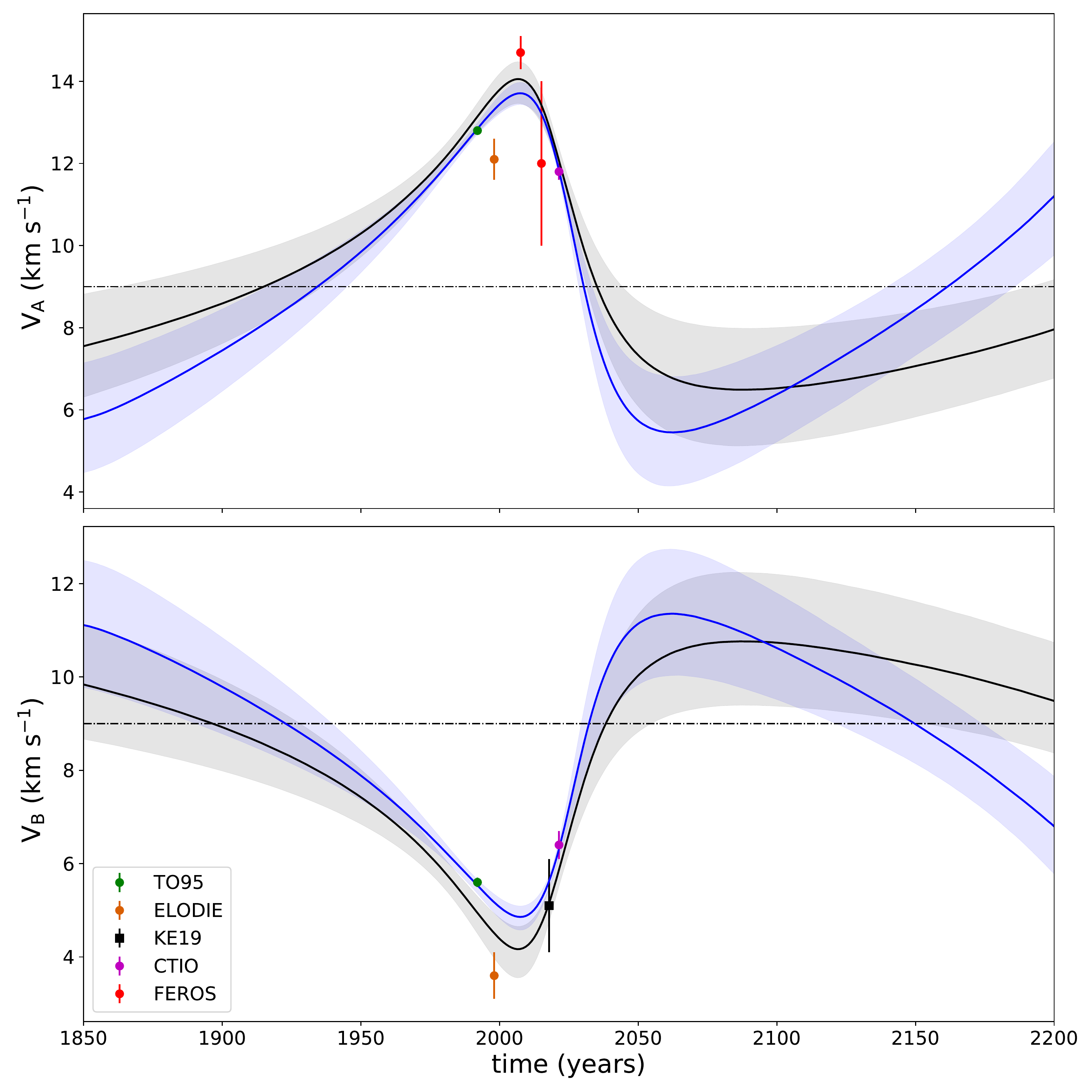}
   \vspace{-0.3cm}
   \caption{Best orbital solution for AB outer orbit for both uncertainty assumptions in the astrometry before 1991. In both panels the solid line corresponds to the best fit model and the shaded area to the $1\sigma$ region. The solid black lines correspond to the best orbital solution assuming small uncertainties ($\sigma\sim0.02$\arcsec) and the blue ones assume large uncertainties ($\sigma\sim0.1$\arcsec).  The dots markers correspond to the RV measurement of systemic velocities from our orbital solutions and the one obtained from CO modelling.}
  \label{Fig:AB_orbit_RV}%
    \end{figure}

All the orbital parameters were estimated from the posterior distributions, taking the median values as the best-fit results and the maximum values between the 16th and 84th percentile as uncertainties (see Table \ref{tab:orbit_AB}). The best-fit orbits for solutions I and II are shown in Fig. \ref{Fig:AB_orbit} and Fig. \ref{Fig:AB_orbit_sepPA}, which are in good agreement with the astrometric measurements. We show the phase-folded RV best-fit orbit in Fig. \ref{Fig:AB_orbit_RV}; the narrow 1-sigma region in this figure mainly comes from the constraints imposed by the AB masses and parallax prior distributions. There are likely small instrumental zero-point offsets among the data sets that were used to determine the systemic velocity variation, which are difficult to determine and could bias the outer orbit solution. As a reminder, all these results rely on the masses and parallax estimated in the orbital fitting of the inner subsystems. The parallax and masses derived from BaBb RV semi-amplitudes ($K_{~\mathrm{Ba}}$ and $K_{~\mathrm{Bb}}$) are proportional to $K^2$ and $K^3$, respectively. Thus, a small systematic error in $K_{~\mathrm{Ba}}$ or $K_{~\mathrm{Bb}}$ can bias the masses and parallax results substantially. The RV amplitudes may be biased, especially for the weakest lines of Bb. Therefore, the masses and the parallax of the BaBb pair that mainly rely on the RVs by \cite{Torres1995} should be considered with caution. A small change in the method of splitting the blended spectra can lead to different masses. The posterior samples of the orbital parameters and all prior distributions used in the MCMC model are available in Appendix \ref{sec:apendix_orbits}.

Accurate astrometry of AB reveals a wavy motion caused by the subsystems (wobble); its amplitude gives an independent constraint on the inner mass ratios. Neglecting the smaller wobble of BaBb, we modelled the astrometry of AB by a combination of two Keplerian orbits, with the orbital parameters of AaAb fixed to the values determined above. A simple least-squares fit yielded the AB orbital parameters similar to solution I, for example $P = 233 \pm 41$ yr. The ratio of the wobble amplitude to the semi-major axis of AaAb was found to be $f = 0.18 \pm 0.04$. Neglecting the influence of the faint light of Ab on the photo-centre of A, this factor gives the inner mass ratio $ q = f/(1-f)=0.22$, compatible within errors with the mass ratio of 0.31 estimated above from the orbit of AaAb.

    \begin{table}[]
    \tiny
                \centering
                \caption{Orbital parameter for HD 98800 AB system.}
                \begin{tabular}{lrrr}
                        \hline
                        \hline  \\
                         & & \multicolumn{2}{c}{This work}  \\     
                          \hline\\  
                        Fitted parameters & KE19 & Solution I\tablefootmark{a} &  Solution II\tablefootmark{b} \\
                          \hline\\
                        Period (years)      &  246 $\pm$ 10  & 230 $\pm$ 20  & 340 $\pm$ 50\\
            T$_{0}$ (years)      &  2023.0 $\pm$ 0.5 &  2023 $\pm$ 1  & 2018 $\pm$ 1\\
            $e$                   &  0.517 $\pm$ 0.007  &  0.46 $\pm$ 0.02 & 0.55 $\pm$ 0.04 \\
            $\omega_{A}$ ($^\circ$)&  63 $\pm$ 2    &  65 $\pm$ 5 &  44 $\pm$ 4 \\
            $\Omega$ ($^\circ$)     &  184.6 $\pm$ 0.2 & 184.5 $\pm$ 0.1 & 184.6 $\pm$ 0.1\\
            $i$ ($^\circ$)      &  88.6 $\pm$ 0.1 &  88.1 $\pm$ 0.1 & 88.39 $\pm$ 0.09 \\
            $\gamma_{~\mathrm{AB}}$ (km s$^{-1}$) & \dots& 8.7 $\pm$ 0.7 & 8.7 $\pm$ 0.9 \\
            $M_{A}$ (M$_{\sun}$) & 1.3 $\pm$ 0.1 &  1.1 $\pm$ 0.3  & 1.2 $\pm$ 0.4 \\
            $M_{B}$ (M$_{\sun}$)  & [1.28] & 1.4 $\pm$ 0.3 &  1.4$\pm$0.4 \\
             $\pi$ (mas)            & [22.2] & 22.2 $\pm$ 0.5  & 22.5 $\pm$ 0.6 \\
            && &  \\
            \hline
            \hline \\
            Derived parameters & & & \\
            \hline\\ 
            $K_{A}$ (km s$^{-1}$) & \dots & 4.2 $\pm$ 0.8 & 3.8 $\pm$ 0.9 \\
            $K_{B}$ (km s$^{-1}$) & \dots & 3.2 $\pm$ 0.8 &  3 $\pm$ 1 \\
            $a$ (") & 1.2 $\pm$ 0.03 & 1.13 $\pm$ 0.08 &  1.5 $\pm$ 0.2 \\
            a (AU)  & \dots &  51 $\pm$ 3  &  67 $\pm$ 8 \\
            d (pc)  & \dots & 45 $\pm$ 1 & 44 $\pm$ 1 \\
                        \hline
                \end{tabular}
                \tablefoot{KE19: \cite{Kennedy2019}. \tablefoottext{a}{Assuming large uncertainties in astrometry before 1991.}\tablefoottext{b}{Assuming small uncertainties in astrometry before 1991.}}
                \label{tab:orbit_AB}
        \end{table}

\section{Short- and long-term future of the quadruple system}\label{sec:N-body}

Several studies investigated the stability of the system over time (e.g. \citealp{Verrier2008}; \citealp{Kennedy2019}), but those studies mostly focused on the stability of the disk around BaBb and less about the evolution of the quadruple system itself. In this section we intend to study both the short- and long-term dynamical evolution of the two pairs of binary systems. Using the new (or revised) orbits obtained for AaAb, BaBb, and AB, we first quantify the dynamical stability over time of the four stars, and second make preliminary predictions for the transit of BaBb and its disk in front of AaAb. To make such predictions, we use the N-body code \texttt{REBOUND}\footnote{Available at \url{https://github.com/hannorein/rebound}} (\citealp{Rein2012}).
For the simulations, we used the orbital solutions for AaAb and BaBb listed in Tables\,\ref{tab:orbit_AaAb} and \ref{tab:orbit_BaBb}, and we tested both solutions I and II for the orbit of AB (Table\,\ref{tab:orbit_AB}). The best-fit parameters are directly taken from the posterior distributions, as their median values.

\subsection{Dynamical stability}

For our dynamical stability analysis, we use the `mean exponential growth of nearby orbits' (MEGNO) criterion implemented in the \texttt{REBOUND} package. As discussed in \citet{Hinse2010}, the MEGNO factor, first introduced in \cite{Cincotta}, provides an estimate of how ordered or chaotic a system is. The MEGNO factor is the integral of variational vectors for a given integration time and a given set of parameters. It is therefore necessary to sample different timescales as the MEGNO is expected to vary over time (and converge to a value of $2$ for a stable system), tracing the different orbital timescales. 
In our case, we want to study the stability of the orbits by changing the masses of Aa and Ab. To do so, we computed the MEGNO value for a matrix of masses, the rows of the matrix consist of $12$ linearly spaced masses for $M_\mathrm{Aa}$ in the range $[0.5,1.5]$\,$\rm{M}_{\odot}$ and $10$ linearly spaced masses for $M_\mathrm{Ab}$ in the range $[0.1,1.0]$\,$\rm{M}_{\odot}$.

\begin{figure}
\centering
\includegraphics[width=0.47\textwidth]{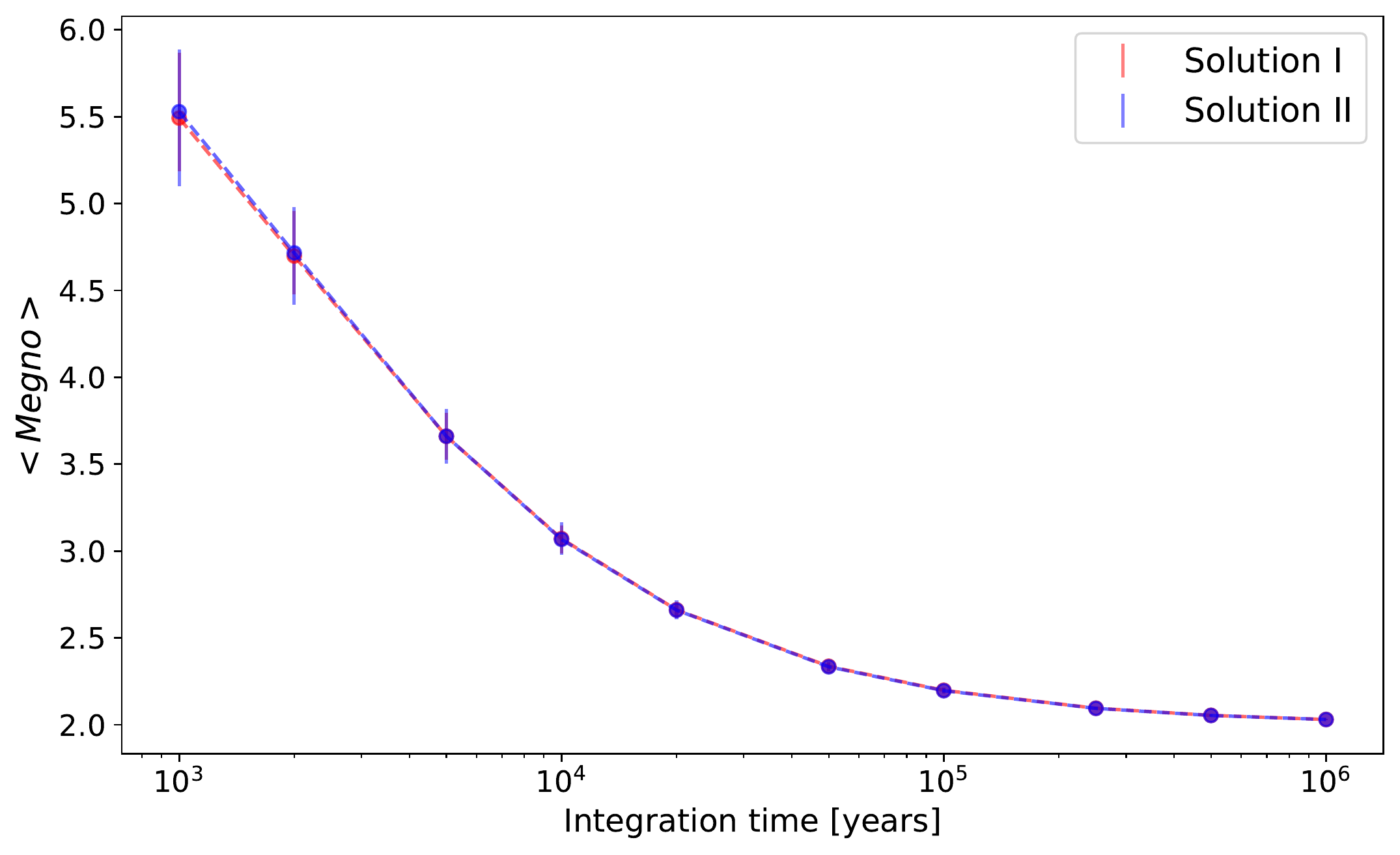}
\caption{Mean MEGNO value for the 12x10 matrix for different integration times. Error bars correspond to $1\sigma$.}
\label{Fig:megno}%
\end{figure}

To setup the simulation, we sequentially added Aa and Ab, and then included a third particle representing BaBb as a single star.
We then integrated the motion of all stars forward in time, using the IAS15 integrator \citep{IAS15}.
Since it is necessary to capture the different timescales for the evolution of the system, we used several integration times, in years: $1\,000$, $2\,000$, $5\,000$, $10\,000$, $20\,000$, $50\,000$, $100\,000$, $250\,000$ $500\,000$, and $1\,000\,000$.
For all the simulations, none of the stars escaped the system, suggesting that it is stable over thousands of AB orbits. Since the final matrices all are homogeneous, in Figure\,\ref{Fig:megno} we show the mean MEGNO value (over the $12\times 10$ matrix) as a function of the integration time, for both solutions I and II, and we computed the standard deviation as the uncertainties. The stability criterion displays an exponential decay, converging towards a value of $2$ (\citealp{Hinse2010}), therefore indicating that the system should be stable over a long period of time, regardless of the uncertainties on the AB orbital parameters. In the exercise above, we treated BaBb as a single star, and it might be worth re-visiting the stability of the system when considering all four stars, but given the large uncertainties on the AB orbit, this is out of the scope of this study.

\subsection{The transit of the disk in front of AaAb}

The orbital parameters of AB strongly suggest that the BaBb pair and its disk will pass in front of the AaAb system \citep{Kennedy2019}, starting sometime in 2026 (depending on the solution used for AB). This presents a unique opportunity to observe and characterise the properties of the dust and gas disk via photometric (and spectroscopic) monitoring of the whole system. In virtue of this possible occultation, we investigate how the photometric light curve might look, including the four stars in the simulation to account for possible interactions between the two binary systems (in App.\,\ref{sec:apendix_nbody} we provide a more detailed explanation on how the simulation is initialised).

To make the predictions for the transit, the starting time of the simulation was set to 2015.17. The choice of the starting date does not matter as we are using the orbital solutions determined in this paper. We integrated the simulation over $18$\,years and saved $10\,000$ intermediate steps (one every $\sim 0.7$ days), saving the positions of the four stars in the reference system centred at the centre of mass of BaBb. The integration was done using the IAS15 integrator, but we also compared our results with the WHfast integrator \citep{WHfast} with a timestep of $0.011$ days, and found no significant differences between the two simulations. Additionally, saving more intermediate steps does not lead to a significant improvement of the resolution of the simulated light curve.

With the $(x, y)$ positions of Aa and Ab, on the plane of the sky, we then estimated if they overlap with the position of the disk, which is centred at the centre of mass of the B system (Fig.\,\ref{Fig:TransitOrbit}). We used the parameters reported in \citet{Kennedy2019}, namely, the inner and outer radii ($2.5$ and $4.6$\,au, respectively), eccentricity ($0.03$), position angle ($15.6^{\circ}$), inclination ($26^{\circ}$), and argument of periapsis ($-73^{\circ}$). To estimate the extinction caused by the circumbinary disk, we first needed the flux ratio between Aa and Ab, and an analytical form for the integrated vertical optical depth of the disk. For the flux ratio, we used the results from the modelling of the PIONIER observations, and the normalised fluxes are $F_\mathrm{Aa} = 0.87$ and $F_\mathrm{Ab} = 0.13$ in the H band. For the vertical optical depth, it is parametrised as $\tau(r) = 0.5 \times r_0/r$, where $r_0$ is the inner radius of the disk ($\tau(r) = 0$ inside and outside the disk). Before applying the extinction law, we first needed to estimate the distance $r$ in the midplane of the disk, accounting for projection and rotation effects. We therefore defined a rotation matrix $\mathcal{R}$ based on the inclination, argument of periapsis, and position angle of the disk, and de-projected the on-sky $(x,y)$ positions of the disk and Aa and Ab stars. The normalised flux at each time-step is then $F_\mathrm{Aa}e^{-\tau(r_\mathrm{Aa})} + F_\mathrm{Ab}e^{-\tau(r_\mathrm{Ab})}$ (the contribution of BaBb is neglected here, but since the vertical optical depth of the disk remains unknown the absolute depth of the transit cannot be constrained).

\begin{figure}
\centering
\includegraphics[width=0.47\textwidth]{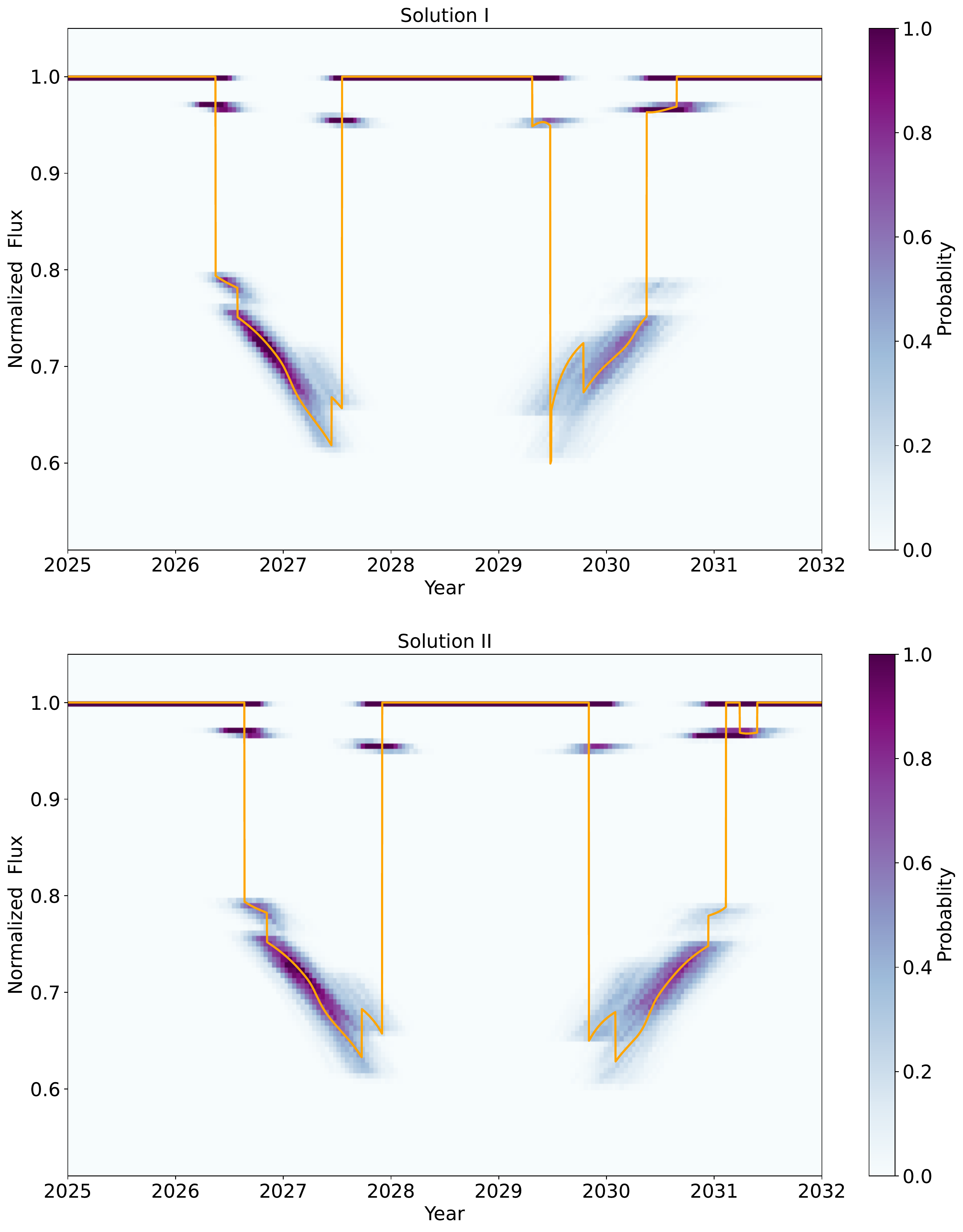}
\caption{Probability density plot of 1000 realizations of the light curve for the occultation of AaAb behind the disk surrounding BaBb for solutions I and II (top and bottom, respectively). The colour bar shows the probability of getting a determined flux at a given time, such that the sum along each of the columns is normalised to unity. In orange we show the light curve for the best-fit parameters (Table\,\ref{tab:orbit_AB}).}
\label{Fig:Lightcurve}%
\end{figure}

We then simulated $1\,000$ transits and their respective light curves by modifying the AB orbital parameters (for both solutions I and II), the parallax, and all four masses randomly drawing $1\,000$ realisations from the MCMC fitting of the AB orbit. This ensures that the correlations between the different parameters are preserved (to avoid, for instance, a small semi-major axis and large stellar masses that would lead to a much shorter orbital period).
Finally, from these $1\,000$ light curves, we estimated a probability distribution of the normalised flux as a function of time, which is shown in Fig. \ref{Fig:Lightcurve}, where the transit light curve for the best-fit solution is shown in orange. The figure shows that the transit should be well constrained in time, and we predict it to start in 2026, going out and passing through the inner regions (devoid of dust) before re-entering behind the northern side of the disk. Our simulations suggest that the transit event should finish sometime between 2030 and 2031. The best-fit solution shows the complex structure of the light curve as one of the stars is sometimes not occulted by the disk.
Comparing the light curves for both solutions I and II, we note that transit starts earlier for solution I, but both cases show a similar behaviour.
Overall, regular photometric monitoring of the quadruple system between 2026 and 2031 at different wavelengths would put unique constraints on the vertical optical depth of the circumbinary disk around BaBb, offering the opportunity to directly measure the surface density of the dust and to possibly derive constraints on the typical size of the dust particles.

\section{Discussion}
Here we discuss the implications of our  results in the context of the formation of this quadruple system and its influence on the disk evolution. A further dynamical simulation of this system is beyond the scope of this paper.

\subsection{Comparisons with previous results}
We refined the orbital results from \cite{Boden2005} and resolved the orbit of the AaAb subsystem for the
first time using PIONIER observations (see Table \ref{tab:orbit_BaBb} and \ref{tab:orbit_AaAb}). Using our orbital solution of BaBb, we derived a dynamical parallax of $22.0\,\pm\,0.6$ mas corresponding to a distance of $45\,\pm\,1$ pc. \cite{Boden2005} placed the system at $42.2\,\pm\,4.7$ pc using their orbital solution, and the updated reduction of the Hipparcos  data \citep{Hipparcos2007} measured a parallax of $22\,\pm\,2$\,mas, corresponding to a distance of $45\,\pm\,5$\,pc likely biased by the unresolved A-B components. There are two entries at the Gaia EDR3 catalogue \citep{GaiaEDR32021} at $\sim0.1\arcsec$ and $\sim0.3\arcsec$ from the positions of AaAb and BaBb, respectively, corresponding to the angular distance after correction using Gaia EDR3 proper motion from J2000 to J2016. Additionally, both subsystems were identified in the cross-matched catalogue between Gaia EDR3 and the Tycho-2 merged with the TDSC \cite[\texttt{I/350/tyc2tdsc}, ][]{GaiaEDR3Xmatch2021}. The parallax values from Gaia EDR3 are $20.1\,\pm\,0.3$ mas and $23.7\,\pm\,0.4$ mas for BaBb and AaAb, respectively. However, both measurements have a large re-normalised unit weight error (RUWE) value \cite[][]{Lindegren2018} and then are considered unreliable. The RUWE value is expected to be around $1.0$ for a good fit to the astrometric observations, while in this case it is $\sim9$ and $\sim6$ for AaAb and BaBb, respectively, meaning that in both cases the unresolved companions produce motion in the photo-centre,  so the 5-parameter Gaia astrometric model performs poorly. The distance inferred with our new results remains consistent with \cite{Boden2005} within $2.3 \sigma$ and is compatible with the Hipparcos value. Using the new distance of BaBb and the orbital solution of AaAb, we derived, for the first time, the dynamical masses of the AaAb binary as $M_{Aa}$ = 0.93 $\pm$ 0.09 M$_{\sun}$ and $M_{Ab}$ = 0.29 $\pm$ 0.02 M$_{\sun}$. Using the \cite{Baraffe2015} 10 Myr isochrones and the dynamical masses of AaAb, we estimated an H-band flux ratio of $15.85\%$ and  $15.06\%$ for solar and sub-solar ($\mathrm{[M/H]}=-0.5$) metallicity, respectively (see Appendix \ref{sec:appendix_fratio}). These flux ratio values are compatible with the flux ratio derived with our PIONIER observations ($\sim\,14\%$, see Table \ref{tab:astrometry}).

\cite{Prato2001} compared the stellar properties derived from near- and mid-infrared diffraction-limited imaging with pre-main sequence evolutionary tracks, yielding masses of $M_{Aa}$ = 1.1 $\pm$ 0.1 M$_{\sun}$, $M_{Ba}$ = 0.93 $\pm$ 0.08 M$_{\sun}$, and $M_{Bb}$ = 0.64 $\pm$ 0.1 M$_{\sun}$ and an age of $\sim10$\,Myr. These values are compatible with the dynamical masses derived in this paper within $\sim 1.5\sigma$. On the other hand, the SED models presented in \cite{Boden2005} suggested stellar properties compatible with the ones published in \cite{Prato2001}. However, the predicted masses from evolutionary tracks were significantly higher than the dynamical masses from \cite{Boden2005}. The authors claimed that this discrepancy came from the assumption of solar abundances in the evolutionary models, proposing sub-solar abundances ($\mathrm{[M/H]}=-0.5$) with an age in the range $8-20$\,Myr. Later, \cite{Laskar2009} estimated a metallicity of $\mathrm{[M/H]}=-0.2 \pm 0.1$ using high-resolution echelle spectra. Additionally, they determined the visible-band flux ratio for Bb/Ba to be $0.416\,\pm\,0.005$. This value is compatible with the visible-band flux ratios estimated from \cite{Baraffe2015} isochrones at 10 Myr and our BaBb dynamical masses results of $0.458$ and $0.428$ for solar and sub-solar ($\mathrm{[M/H]}=-0.5$) metallicity, respectively (see Appendix \ref{sec:appendix_fratio}). Given the uncertainty on the derived dynamical masses due to the degeneracy between  $K_{Ba}$ and $K_{Bb}$, we cannot use the quadruple system  yet  to benchmark evolutionary track models, calling for additional observations to better constrain both the orbital solutions and the abundances of the four stars. Both  I and II AB orbital solutions feature comparable values for the inclination and longitude of the ascending node $\Omega$,  within $\lesssim 0.5^\circ$ from the latest orbital solution \cite[][see Table \ref{tab:orbit_AB}]{Kennedy2019}. This result shows that despite the fact that the orbit of AB will remain uncertain for several years as more observations are collected, its orientation is already well constrained and robust.

\subsection{Mutual alignment}

\begin{table}[]
                \centering
                \caption{Mutual inclinations between all orbital planes in HD\,98800 and the circumbinary disk.}
                \begin{tabular}{lrr}
                        \hline
                        \hline  \\
                         Mutual inclination & \multicolumn{2}{c}{} \\
                         \hline\\
                        $\Phi_{BaBb-AB}$ ($^\circ$)   & \multicolumn{2}{c}{146.8 $\pm$ 0.5} \\
            $\Phi_{AaAb-AB}$ ($^\circ$)   & \multicolumn{2}{c}{49.2 $\pm$ 0.1} \\
            $\Phi_{AaAb-BaBb}$ ($^\circ$) & \multicolumn{2}{c}{157.3 $\pm$ 0.5} \\
            \\ 
             & $i_{disk}=26^{\circ}$  & $i_{disk}=154^{\circ}$ \\
                        \cline{2-3}\\ 
                        $\Phi_{BaBb-Disk}$ ($^\circ$) & 89 $\pm$ 1 & 134.2 $\pm$ 1 \\
            $\Phi_{AaAb-Disk}$ ($^\circ$) & 111 $\pm$ 1 &  23 $\pm$ 1 \\
            $\Phi_{AB-Disk}$ ($^\circ$)   & 63 $\pm$ 1 & 66   $\pm$ 1 \\
            \hline \\
                \end{tabular}
                \label{tab:disk_incl}
        \end{table} 
        
The mutual inclinations between the inner and outer orbits in a hierarchical system can constrain the initial conditions of its formation \citep{Fekel1981,Sterzik2002}. Hierarchical fragmentation of a rotating cloud \citep{Bodenheimer1978} or fragmentation of a circumbinary disk \citep{Bonell1994} should result in near co-planar configurations. On the other hand, misaligned orbits could be the result of turbulent fragmentation or dynamical interactions \citep{Lee2019b}. Similarly, the relation between circumbinary disk orientation and the orbital parameters of the host binary can be used to better constrain their formation scenarios \citep{Czekala2019}. The relative inclination $\Phi$ between the inner and outer orbits (or disk) is given by

\begin{eqnarray}
                \cos \Phi &=& \cos\, i_1\, \cos i_2\, + \sin\, i_1\, \sin i_2\, \cos \left(\Omega_1 - \Omega_2 \right), 
                \label{eq:mut_incl}
\end{eqnarray}

\noindent where $i_{1},\,i_{2}$ are the inclinations of each orbit (or disk and orbit) and $\Omega_{1},\,\Omega_{2}$ are the corresponding longitudes of the ascending nodes. The mutual inclination $\Phi$ ranges from $0^{\circ}$ to $180^{\circ}$, where $\Phi=0^{\circ}$ corresponds to co-planar and co-rotating orbits. When $\Phi>90^{\circ}$ the systems are retrograde, and $\Phi=90^{\circ}$ means polar configuration. The circumbinary disk of BaBb was initially thought to be co-planar with the host binary \citep{Tokovinin1999,Prato2001}, but recent ALMA observations revealed that the circumbinary disk is actually in polar configuration \citep{Kennedy2019}. Additionally, \cite{Giuppone2019} suggested that the near polar configuration between the circumbinary disk and BaBb orbit is the most stable configuration among all possible disk inclinations. Given that we reduced the uncertainty of $i$ and $\Omega$ for the BaBb orbit from $\sim 3^{\circ}$ to $\sim 0.5^{\circ}$, it is important to re-calculate the mutual inclination. \cite{Kennedy2019} found that the disk is inclined either by $26^{\circ}$ or $154^{\circ}$ with respect to the sky plane. The $\Omega_{disk}$ published in \cite{Kennedy2019} is defined as the node where rotation of the disk is moving towards the observer, that is with a difference of $\pi$ with respect to our convention. For consistency, we added $\pi$ to the published value resulting in $\Omega_{disk}=196^{\circ} \pm 1^{\circ}$. The new mutual inclinations between all the orbital planes of HD\,98800 are reported in Table \ref{tab:disk_incl}, including the mutual inclination of the disk with respect to the inner and outer binaries. For these mutual inclination values, we used the AB orbital parameters from solution I (see Table \ref{tab:orbit_AB}). Given that the inclination and $\Omega$ value of solutions I and II are close to each other within $\sim 0.1^\circ$, the subsequent analysis  remains valid for both outer orbit solutions. The uncertainties were calculated using a Monte-Carlo uncertainty propagation, assuming Gaussian errors. We confirmed the near polar configuration of the disk relative to the orbital plane of BaBb in the case $i_{disk}=26^{\circ}$ and found $\Phi_{BaBb-Disk} = 134.2^{\circ}$ in the case $i_{disk}=154^{\circ}$. Using the posterior distributions from our fitting, and the posteriors from the disk fitting from \citet{Kennedy2019}, yields a nominally significant misalignment of the disk angular momentum vector and the binary pericentre vector; $1.7 \pm 0.5^\circ$. In principle, this misalignment provides a measurement of the disk mass, but given likely systematic uncertainties, for example, in estimating blended RVs, we consider this measurement to be an upper limit. Updating the calculation from \citet{Martin2019} using the 99.7th percentile from our posteriors, the upper limit on the disk mass is 0.02\,$M_\odot$. The angle from polar is slightly smaller, but the binary mass is larger, so our limit is essentially the same as the upper limit computed by \citeauthor{Martin2019}. Circumbinary disks are preferentially co-planar around short period ($<40$ days) host binaries \citep{Czekala2019}, while for longer orbital periods, mutual inclinations are found in a wide range of configurations \citep{Kennedy2019,Gravity2021,Czekala2021}. In that regard, determining the orbital parameters of binaries and the mutual inclination of the circumbinary disk at intermediate periods (40 - 300 days), such as the presented HD\,98800 results, can contribute to better understand the dynamical scenario leading to co-planar or polar disk configurations.

\subsection{Formation history of HD 98800}

The HD\,98800 system is a member of the TWA Hydrae young loose association \citep{Torres2008}, therefore it is unlikely that it experienced strong external dynamical interactions with other stars. In general, hierarchical systems that formed under high dynamical interactions between nascent protostars have misaligned and eccentric orbits, and their masses are not comparable \citep{Sterzik2002}.  The AB and AaAb orbits are moderately misaligned (see Table \ref{tab:disk_incl}) and, excluding the Ab component, the masses are comparable. It is expected that the orbital and physical parameters of this quadruple system contain imprints of its formation scenario. Near co-planarity and comparable masses  in wide solar-type hierarchical systems can be a sign of their formation from a common core \citep{Tokovinin2020Jun}. The collapse of two nearby clouds and their inward accretion-driven migration by accretion \citep{TokovininMoe2020} can result in compact co-planar hierarchical systems with moderate eccentricities and period ratios. However, HD\,98800  is a quadruple system with a 2+2 configuration where the inner orbits are counter-rotating and the BaBb is misaligned with the outer orbit AB.

The encounter of two clumps can create shock fronts that lead to the fragmentation of each core into a binary, forming a 2+2 quadruple system \citep{Whitworth2001}. Hypothetically, this formation scenario can produce wide quadruple systems with similar masses between all four components and comparable inner periods, called $\epsilon$\,Lyr type \citep{Tokovinin2008}, where the inner orbits are expected to be mutually misaligned. Generally, $\epsilon$\,Lyr type have wide outer separations ($P_{outer}\gtrsim 450$ kyr), but more compact 2+2 systems are known as well (HIP\,41171, $P_{outer}\sim900$ yr, \citealt{Tokovinin2019b}: FIN\,332, $P_{outer}\sim3000$ yr, \citealt{Tokovinin2020Sep}). Although the outer period of HD\,98800 is shorter than usual for these systems ($P\sim 200-400$ yr), the orbital configuration still matches this $\epsilon$\,Lyr type except for the expectation of similar masses of its components. The mass-ratio of BaBb and AB are $\sim0.8$ and $\sim0.9$, respectively, while the mass-ratio of AaAb binary is $\sim0.31$.

The large BaBb eccentricity and its counter-rotating configuration with respect to the AaAb and AB orbits could be explained as the result of dynamical interactions. Tidal forces may have ripped away circumbinary material from AaAb, and in the same way, may have perturbed the BaBb circumbinary disk and the eccentricity of the host binary. In consequence, the formation process of the HD\,98800 system remains unclear.
 
\subsection{The low mass ratio of AaAb and its lack of a disk}

An intriguing characteristic of HD 98800 is that it still holds a massive circumbinary gas disk around the system BaBb \citep{Ribas2018,Kennedy2019}. Nonetheless, no circumbinary disk has been found around the system AaAb. A possible explanation for the persistent existence of the detected disk has been proposed by \citet{Ribas2018}. These authors speculated that the disk has survived for so long because of the tidal torques exerted by BaBb on the inner edge and by AB on the outer edge, which stopped or significantly reduced viscous accretion, leading to a scenario in which the disk is only losing mass due to photo-evaporation. On the other hand, the lack of a disk around system A, which could have evolved in a similar way as the disk around B, could be related to a faster disk dispersal due to a higher X-ray luminosity, estimated to be $\sim$ 4 times the one of system B \citep{Kastner2004}.

Recently, with a 1D+1D model of gas disk evolution, \citet{Ronco2021} explored the scenario proposed by \citet{Ribas2018} in arbitrary hierarchical triple star systems and, particularly, in HD\,98800. They show that the current age and mass of gas of HD\,98800 B can be reproduced if the disk was originally an intermediate to high-mass disk ($\sim0.05 - 0.1\,\mathrm{M_\odot}$), and if it had a moderate to low viscosity ($10^{-4}-10^{-3}$). 
To evaluate the current non-existence of a disk around system A, these authors considered, for simplicity, that both the disk parameters and the characteristics of system A (i.e. its mass ratio and separation) were the same as those of the system B, except for its higher X-ray luminosity, as suggested by \citet{Kastner2004}. Under these considerations, their simulations show that the possible disk around A may have dissipated in less than $7-10$\,Myr, the estimated age of HD\,98800.  We know that the assumption of equal inner mass ratios in HD\,98800 does not hold. However, \citet{Ronco2021} also show that the smaller the mass ratio of the inner binary in a hierarchical triple star system, the faster the circumbinary disk dissipates, suggesting that the disk around system A in HD\,98800 may have dissipated even faster. Our new findings and the characterisation of system A,  presented in sec. \ref{sec:AaAb-Orbit}, effectively show a mass ratio that is much lower than that of system B, reaffirming this possibility and contributing to the explanation of the absence of the A disk.

\section{Summary and conclusions}

In this work, we present a new orbital solution for the HD\,98800 quadruple system. Using PIONIER observations, we obtained new astrometric positions and a flux ratio of AaAb and BaBb subsystems. We refined the orbital solution presented by \cite{Boden2005} and derived, for the first time, the full orbital solution for the AaAb binary. From our orbital solution, we confirmed the polar configuration of the circumbinary disk around BaBb. Using the dynamical parallax of BaBb, we calculated the dynamical masses of the AaAb pair. The dynamical masses and parallax are strongly dependent on the RV semi-amplitude $K_{Ba}$ and $K_{Bb}$, estimated mainly from the RV measurements by \cite{Torres1995}. New high-resolution spectroscopic observations of HD\,98800 could remove possible biases in the estimation of the RV semi-amplitude of the inner systems. Spectroscopic observation with adaptive optics correction could allow us to acquire resolved spectra of each subsystem, thus avoiding line blending of the four components. The estimated visible-band AaAb flux ratio is $\lesssim 1 \%$ \citep{Laskar2009}, making it difficult to disentangle the RV of Ab. From our PIONIER observations, we estimated an H-band flux ratio of $\sim14\%$ for the AaAb binary. This more favourable flux ratio opens the possibility to measure, for the first time, the RV of Ab using high resolution infrared spectroscopy. This would allow us to calculate the dynamical masses and parallax of Aa and Ab independently from the parallax of BaBb. Spectroscopic monitoring of HD\,98800 is relevant not only for more robust dynamical masses and parallax estimates, but also to properly establish the abundances of the four stars. These measurements will provide valuable inputs to test and improve pre-main sequence evolutionary models and better constrain models of dust disk evolution.

We tested the dynamical stability of the quadruple using N-body simulations. Using the orbital parameters and the mass values of the inner binaries, the simulation probed the long-term stability of this system for both outer orbit solutions; we found that the system should be stable over thousands of orbital periods. The AB outer orbit predicts that the AaAb binary will pass behind the disk around BaBb in the coming years. Using our N-body simulation, we predicted that the transit will start in 2026 and should finish between 2030-2031. This transit presents an unprecedented opportunity to characterise the disk structure along a $\sim10$\,au long chord, with the width of this chord set by the projected extent of the AaAb orbit.

From mass ratios, periods, eccentricities, and mutual orbit orientations, we evaluate possible formation scenarios for HD\,98800. The similarity of the components' masses suggests a common formation history. The misalignment between the orbital planes of the inner binaries and the high eccentricity of the BaBb pair suggest a possible dynamical perturbation.  Assuming AaAb as a binary exactly equal to that of system B, but with a higher X-ray luminosity as suggested by \citet{Kastner2004},
simulations from \cite{Ronco2021} show that the disk around A can dissipate in less than 10\,Myr due to photo-evaporation. This scenario can explain the lack of a circumbinary disk around the AaAb subsystem. These authors also show that a lower mass ratio could indeed promote  faster photo-evaporation of the disk. Thus, the low mass ratio derived here actually agrees with faster disk dispersal.

With the current observational evidence, we cannot properly establish the formation process of HD\,98800 as there are still some uncertainties in the parallax of A as well as in the orbit of AB. Recently, other works have also used long-baseline infrared interferometry to characterise hierarchical multiple systems \citep{Kraus2020,Gravity2021,Czekala2021}. Further monitoring of other hierarchical systems, especially at young ages (1-100 Myr), in combination with large survey data, will improve our understanding of the formation and dynamical evolution of these kinds of systems.  

\begin{acknowledgements}
      The authors would like to thank the anonymous referee for constructive comments that helped to improve the content and clarity of this paper. S.Z-F acknowledges financial support from the European Southern Observatory via its studentship program and ANID via PFCHA/Doctorado Nacional/2018-21181044. J.O. acknowledges support from the Universidad de Valpara\'iso and from Fondecyt (grant 1180395). S.Z-F., A.B., J.O. and M.P.R acknowledge support by ANID, -- Millennium Science Initiative Program -- NCN19\_171. A.B acknowledges support from Fondecyt (grant 1190748). M.P.R. acknowledges support from Fondecyt (grant 3190336). G.M.K. is supported by the Royal Society as a Royal Society University Research Fellow. This research made use of \textsf{exoplanet} \citep{exoplanet:exoplanet} and its dependencies \citep{exoplanet:astropy13, exoplanet:astropy18, exoplanet:pymc3, exoplanet:theano}. This research has made use of the Washington Double Star Catalog maintained at the U.S. Naval Observatory. This work has made use of data from the European Space Agency (ESA) mission {\it Gaia} (\url{https://www.cosmos.esa.int/gaia}), processed by the {\it Gaia} Data Processing and Analysis Consortium (DPAC, \url{https://www.cosmos.esa.int/web/gaia/dpac/consortium}). Funding for the DPAC has been provided by national institutions, in particular the institutions participating in the {\it Gaia} Multilateral Agreement. This publication makes use of VOSA \citep{Bayo2008}, developed under the Spanish Virtual Observatory project supported by the Spanish MINECO through grant AyA2017-84089. This research has made use of NASA’s Astrophysics Data System.
\end{acknowledgements}

%
%
\bibliography{biblio}

\begin{appendix}
\section{Observations}\label{sec:apendix_observations}

This section presents complementary information regarding the observations used in this work. The calibrator stars used in our PIONIER observations are listed in Table \ref{tab:calibrators}. These calibrators were chosen using the \texttt{SearchCal} tool.

\begin{table}[h]
\tiny
                \centering
                \caption{Calibrator stars for HD 98800 observations. The distance column refers to the calibrator to science object angular distance in degrees.}
                \begin{tabular}{lccr}
                        \hline
                        \hline  \\
                        SIMBAD id & distance (deg) & V mag & H mag \\
                          \hline\\
                        HD 98828  & 0.42 & 7.83  & 5.35 \\
            HD 98729  & 0.79 & 7.77 & 5.42 \\
            & &  &\\
            \hline
            \hline 
                \end{tabular}
                \label{tab:calibrators}
        \end{table}

Most of the RV measurements used in this work were published by \cite{Torres1995}. Here we present the new RV measurements from CHIRON observations and science ready archive spectra, see Table \ref{tab:appendix_RV_AaAb} and Table \ref{tab:appendix_RV_BaBb}. The RVs used in the orbital fitting of the AB orbit are listed in Table \ref{tab:appendix_RV_AB}.

\begin{table}[h]
\tiny
                \centering
                \caption{Radial velocity measurements for AaAb subsystem.}
                \begin{tabular}{lcccr}
                        \hline
                        \hline  \\
                        MJD &  $~\mathrm{RV_{Aa}}$  & $~\mathrm{\sigma_{Aa}}$ & (O-C) & Instrument \\
                        & (km s$^{-1}$) & (km s$^{-1}$) & (km s$^{-1}$) &\\
                          \hline\\
                        50841.1287 & 20.1031 & 0.5 & 0.0805 & ELODIE \\
            50842.1262 & 19.9605 & 0.5 & -0.0375 & ELODIE \\
            54308.4955 & 12.156 & 0.8 & -0.5271 & FEROS \\
            54309.4913 & 12.218 & 0.8 & -0.1113 & FEROS \\
            54309.5368 & 12.247 & 0.8 & -0.0667 & FEROS \\
            54310.4710 & 12.036 & 0.8 & 0.0315 & FEROS \\
            54311.4629 & 11.713 & 0.8 & 0.0134 & FEROS \\
            54312.4900 & 11.572 & 0.8 &  0.1637 & FEROS \\
            54314.4636 & 11.270 & 0.8 & 0.3575 & FEROS \\
            54315.4672 & 11.022 & 0.8 & 0.3306 & FEROS \\
            57062.2727 & 9.919 & 0.8 & 0.2176 & FEROS \\
            59323.1553 & 15.05 & 0.5 & -0.2036 & CHIRON \\
            59338.0931 & 12.26 & 2.5 & 3.4653 & CHIRON \\
            59411.9949 & 7.213 & 0.5 & -0.9481 & CHIRON \\
            59421.9871 & 8.869 & 0.2 & 0.1754 & CHIRON \\
            59424.9528 & 8.899 & 0.2 &  0.0446 & CHIRON \\
            & &  & &\\
            \hline
            \hline \\
                \end{tabular}
                \label{tab:appendix_RV_AaAb}
        \end{table}

        \begin{table}[h]
\tiny
                \centering
                \caption{Radial velocity measurements for BaBb subsystem.}
                \begin{tabular}{lcccr}
                        \hline
                        \hline  \\
                        MJD &  $~\mathrm{RV_{Ba}}$  & $~\mathrm{\sigma_{Ba}}$ & (O-C) & Instrument \\
                        & (km s$^{-1}$) & (km s$^{-1}$) & (km s$^{-1}$) &\\
                          \hline\\
                        50841.1287 & 8.81 & 2.7 & -3.8347 & ELODIE \\
            50842.1262 & 8.35 & 2.7  & -4.4114 & ELODIE \\
            59323.1553 & 14.029 & 1.0 & 0.4724 & CHIRON \\
            59421.9871 & -20.740 & 0.2 & 0.0881 & CHIRON \\
            59424.9528 & -19.045 & 0.2 & -0.2985 & CHIRON \\
            \hline      \\
                        MJD &  $~\mathrm{RV_{Bb}}$  & $~\mathrm{\sigma_{Bb}}$ & (O-C) & Instrument \\
                        & (km s$^{-1}$) & (km s$^{-1}$) & (km s$^{-1}$) &\\
                         \hline\\
            50841.1287 & -7.40 & 0.7 & 0.6603 & ELODIE \\
            50842.1262 & -7.82 & 0.7  & 0.3855 & ELODIE \\
            59323.1553 & -2.752 & 1.5 & -0.2293 & CHIRON \\
            59421.9871 & 40.533 & 0.5 & 0.2658 & CHIRON \\
            59424.9528 & 38.587 & 0.5 & 0.9122 & CHIRON \\
            & &  & &\\
            \hline
            \hline \\
                \end{tabular}
                \label{tab:appendix_RV_BaBb}
        \end{table}

                \begin{table}[h]
\tiny
                \centering
                \caption{Radial velocity measurements for AB system.}
                \begin{tabular}{lccrr}
                        \hline
                        \hline  \\
                        Median MJD &  $~\mathrm{RV_{A}}$ & $~\mathrm{\sigma_{A}}$ & (O-C)\tablefootmark{a} & Source \\
                        & (km s$^{-1}$) & (km s$^{-1}$) & (km s$^{-1}$)  & \\
                          \hline\\
            48635.4564 & 12.8 & 0.1 & 0.0533 & TO95 \\
            50841.6274 & 12.1 & 0.5 & 1.2209 & ELODIE \\
            54311.9669 & 14.7 & 0.4 & -0.9789 & FEROS \tablefootmark{b}\\
            57062.7727 & 12 & 2 & 1.2147 & FEROS \tablefootmark{c}\\
            59375.5439 & 11.8 & 0.2 & 0.0359 & CTIO \\
            \hline\\
            Median MJD &  $~\mathrm{RV_{B}}$ & $~\mathrm{\sigma_{B}}$ & (O-C)\tablefootmark{a} & Source \\
                        & (km s$^{-1}$) & (km s$^{-1}$) & (km s$^{-1}$)  & \\
                        \hline\\
                        48635.4564 &  5.6 & 0.1 & -0.0682 & TO95 \\
            50841.6274 & 3.4 & 0.7 & 1.5715 & ELODIE \\
            58072.3724 & 5.1 & 1 & 0.5269 & KE19 \tablefootmark{d} \\
            59375.5439 & 6.4 & 0.4 & -0.0884 & CTIO \\
            & & & & \\
            \hline
            \hline 
                \end{tabular}
                \tablefoot{\tablefoottext{a}{(O-C) from solution I.}\tablefoottext{b}{From FEROS observations taken in 2007.}\tablefoottext{c}{From FEROS observation taken in 2015.}\tablefoottext{d}{\cite{Kennedy2019}.}}
                \label{tab:appendix_RV_AB}
        \end{table}

The AB astrometric measurements before 2016 are available at the Washington Double Star catalogue \citep[WDS,][]{WDS2001} and \cite{Tokovinin2018a}. The new astrometry measurement from speckle interferometry at SOAR are listed in Table \ref{tab:appendix_ABastrometry}.

\begin{table}[h]
\tiny
                \centering
                \caption{Astrometry measurements of AB system.}
                \begin{tabular}{lrccrcc}
                        \hline
                        \hline  \\
                        Date &  sep  & $\sigma_{\mathrm{sep}}$ & (O-C)$_{\mathrm{sep}} \tablefootmark{a}$ & P.A.  & $\sigma_{\mathrm{PA}}$ & (O-C)$_{\mathrm{P.A.}} \tablefootmark{a}$ \\
                        & (\arcsec) & (\arcsec) & (\arcsec) & ($^\circ$) & ($^\circ$) & ($^\circ$) \\
                          \hline\\
                        1909.5 & 1.0 & 0.1 & 0.3106 & 190.0 & 2.0 & -1.2605\\
            1910.3 & 0.8 & 0.1 & 0.1243 & 180.0 & 2.0 & 8.8414 \\
            1912.66 & 0.65 & 0.1 & 0.0153 & 187.0 & 2.0 & 2.1753 \\
            1926.25 & 0.41 & 0.1 & 0.0266 & 192.5 & 2.0 & 0.0270 \\
            1930.39 & 0.37 & 0.1 & 0.0667 & 194.0 & 2.0 & 0.6809 \\
            1936.32 & 0.23 & 0.1 & 0.0422 & 204.7 & 2.0 & -3.7228 \\
            1937.83 & 0.24 & 0.1 & 0.0813 & 204.1 & 2.0 & -0.0558 \\
            1937.98 & 0.18 & 0.1 & 0.0241 & 206.8 & 2.0 & -2.3872 \\
            1959.25 & 0.18 & 0.1 & -0.1072 & 348.7 & 1.0 & 6.4093 \\
            1960.27 & 0.2 & 0.1 & -0.1068 & 0.0 & 1.0 & -4.1886 \\
            1963.85 & 0.26 & 0.1 & -0.1151 & 358.7 & 1.0 & -1.0060 \\
            1964.35 & 0.3 & 0.1 & -0.0845 & 1.1 & 1.0 & -3.1961 \\
            1967.28 & 0.32 & 0.1 & -0.1188 & 1.9 & 1.0 & -2.9466 \\
            1976.13 & 0.52 & 0.1 & -0.0713 & 2.2 & 1.0 & -1.1944 \\
            1979.21 & 0.59 & 0.1 & -0.0486 & 1.1 & 1.0 & 0.3996 \\
            1991.25 & 0.775 & 0.01 & -0.0021 & 2.9 & 1.0 & 0.0363 \\
            1991.3882 & 0.777 & 0.01 & -0.0011 & 3.2 & 1.0 & -0.2498 \\
            1996.1826 & 0.807 & 0.02 & 0.0045 & 3.1 & 1.0 & 0.3122 \\
            2004.0860 & 0.78 & 0.01 & -0.0032 & 3.0 & 1.0 &  1.1590 \\
            2006.1913 & 0.745 & 0.01 & -0.0169 & 3.7 & 1.0 & 0.6725 \\
            2009.2638 & 0.7139 & 0.002 & -0.0021 & 4.22 & 0.71 & 0.4920 \\
            2009.2638 & 0.7144 & 0.002 & -0.0016 & 4.22 & 0.9 & 0.4920 \\
            2009.2638 & 0.714 & 0.002 & -0.0020 & 4.26 & 0.31 & 0.4520 \\
            2011.0355 & 0.6853 & 0.002 & 0.0045 & 5.04 & 0.61 & -0.1085 \\
            2011.0355 & 0.6877 & 0.002 & 0.0069 & 4.95 & 0.98 & -0.0185 \\
            2013.1272 & 0.63 & 0.002 & -0.0003 & 5.33 & 0.34 & -0.1047 \\
            2013.1272 & 0.6291 & 0.002 & -0.0012 & 5.34 & 0.56 & -0.1147 \\
            2014.0581 & 0.6021 & 0.002 & -0.0027 & 4.98 & 0.43 & 0.3929 \\  
            2015.1696 & 0.571 & 0.002 & -0.0006 & 5.56 & 0.26 & 0.0073 \\
            2015.1696 & 0.572 & 0.002 & -0.0001 & 5.46 & 0.71 & 0.1073 \\
            2016.0485 & 0.546 & 0.002 & 0.0019 & 5.74 & 0.33 & -0.0015 \\
            2016.9603 & 0.513 & 0.002 & 0.0005 & 5.46 & 0.33 & 0.4765 \\
            2018.0856 & 0.471 & 0.002 & -0.0009 & 6.68 & 0.33 & -0.4625 \\
            2019.1399 & 0.429 & 0.002 & -0.0021 & 6.46 & 0.20 & 0.0700 \\
            2019.9503 & 0.395 & 0.002 & -0.0036 & 6.92 & 0.22 & -0.1055 \\
            2020.9961 & 0.357 & 0.002 & 0.0023 & 7.75 & 0.22 & -0.4891 \\
            2021.3159 & 0.344 & 0.002 & 0.0022 & 7.08 & 0.24 & 0.3404 \\
            & &  & & & & \\
            \hline
            \hline\\
                \end{tabular} 
                \label{tab:appendix_ABastrometry}
                \vspace{-0.5cm}
                \tablefoot{\tablefoottext{a}{(O-C) from solution I.}}
        \end{table}

\section{Orbital fitting complementary information}\label{sec:apendix_orbits}
\begin{figure*}
\centering
\begin{subfigure}{\textwidth}
\centering
\includegraphics[width=0.95\linewidth]{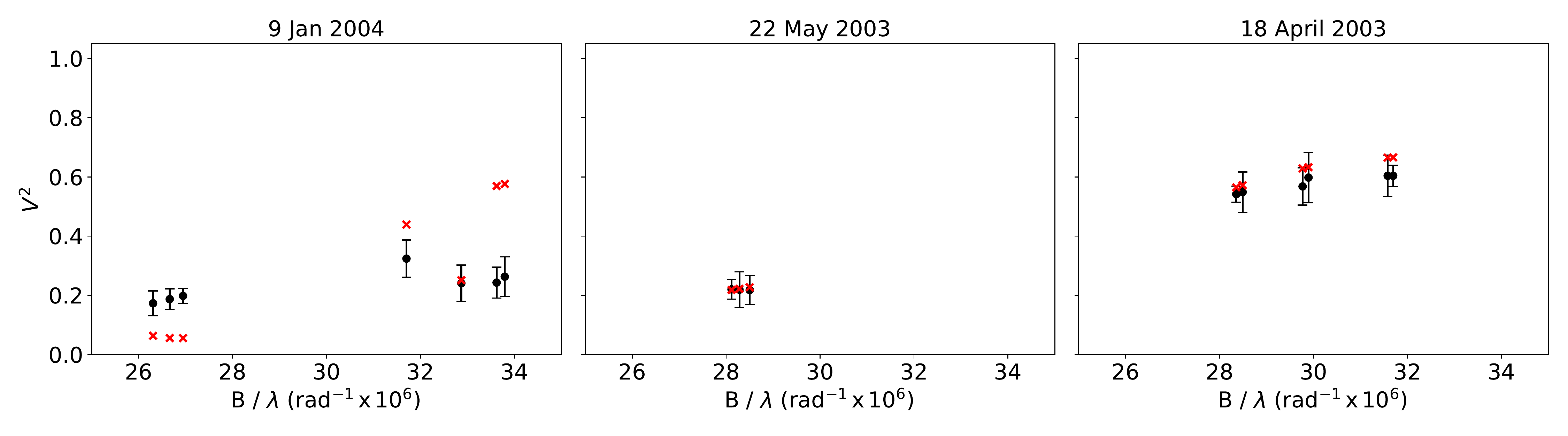}
\end{subfigure}

\begin{subfigure}{\textwidth}
\centering
\includegraphics[width=0.95\linewidth]{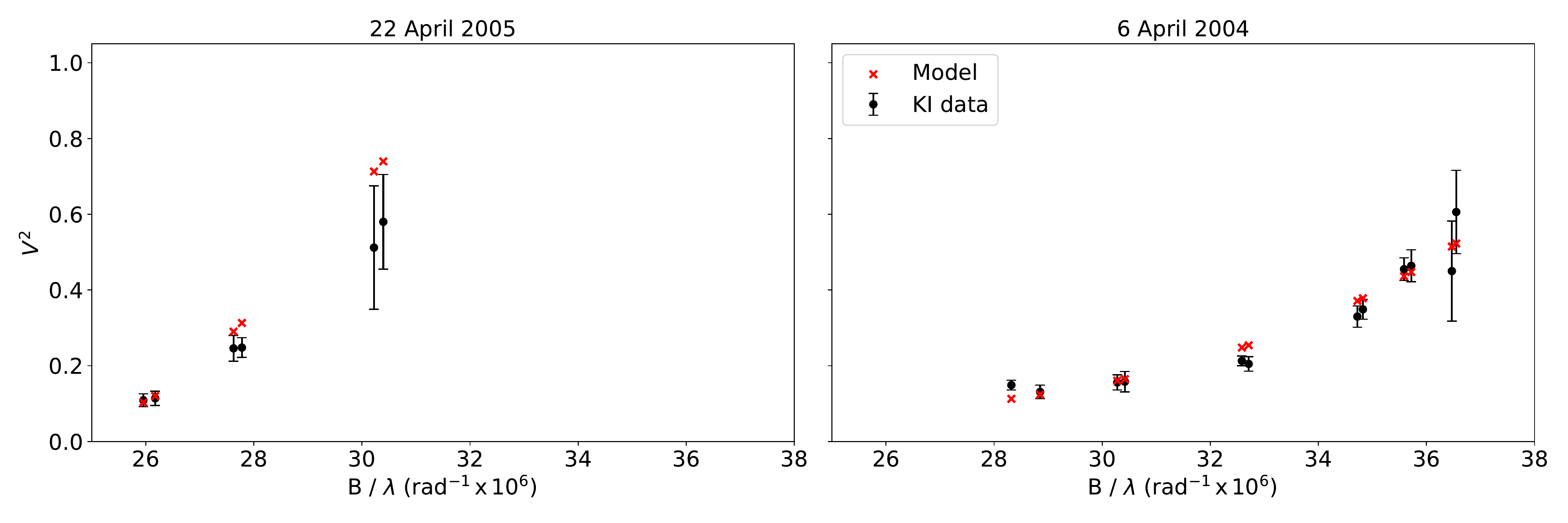}
\end{subfigure}

\caption{Squared visibilities from Keck Interferometer observations published in \cite{Boden2005}. The black circles represent the observed values and the red crosses represent the best-fit BaBb binary model from this work.}
\label{fig:KI_model}
\end{figure*}

This section presents the prior distributions used for each orbital fitting. Additionally, we also show the corner plots from the posterior samples of each MCMC model. Fig. \ref{fig:KI_model} shows the $V^2$ from KI observations and the best fit binary model from the BaBb orbital fitting result. 

\begin{table}[h]
\tiny
                \centering
                \caption{Prior distribution used in AaAb and BaBb orbital fitting.}
                \begin{tabular}{lrr}
                        \hline
                        \hline  \\
                        Parameters &  AaAb & BaBb \\
                          \hline\\
                        Period (days)       & \texttt{LogUniform} $[200,\,300]$  & LogUniform$[250,\,350]$  \\
            T$_{0}$ (MJD)       & \texttt{Normal} $[48\,737,\,20]$ & \texttt{Normal} $[48\,709,\,20]$ \\
            $e$                 & \texttt{Uniform} $[0,\,1]$  & \texttt{Uniform} $[0,\,1]$ \\
            $\omega_{Aa/Ba}$ (rad) & \texttt{Uniform} $[0,\,2\pi]$ & \texttt{Uniform} $[0,\,2\pi]$ \\
            $\Omega$ (rad)      & \texttt{Uniform} $[0,\,2\pi]$  & \texttt{Uniform} $[0,\,2\pi]$ \\
            $~\mathrm{cos\,(i)}$ & \texttt{Uniform} $[-1,\,1]$  & \texttt{Uniform} $[-1,\,1]$  \\
            $a$ (mas)            & \texttt{Uniform} $[5,\,30]$  & \texttt{Uniform} $[5,\,30]$  \\
            K$_{1}$ (km s$^{-1}$)        & \texttt{Uniform} $[0,\,20]$ & \texttt{Uniform} $[0,\,50]$ \\
            K$_{2}$  (km s$^{-1}$)       & \dots  & \texttt{Uniform} $[0,\,50]$  \\
            $\gamma$       & \texttt{Uniform} $[0,\,20]$ & \texttt{Uniform} $[0,\,20]$ \\
            & &  \\
            \hline
            \hline 
                \end{tabular}
                \label{tab:appendix_priorInner}
        \end{table} 
        
\begin{table}[h]
                \centering
                \caption{Prior distribution used in AB orbital fitting.}
                \begin{tabular}{lr}
                        \hline
                        \hline  \\
                        Parameters &  AB\\
                          \hline\\
                        Period (years)       & \texttt{LogUniform} $[100,\,500]$ \\
            T$_{0}$ (years)       & \texttt{Uniform} $[2\,000,\,2\,040]$ \\
            $e$                 & \texttt{Uniform} $[0,\,1]$  \\
            $\omega_{A}$ (rad) & \texttt{Uniform} $[0,\,2\pi]$  \\
            $\Omega$ (rad)      & \texttt{Uniform} $[0,\,2\pi]$   \\
            $~\mathrm{cos\,(i)}$ & \texttt{Uniform} $[-1,\,1]$    \\
            $M_A$ (M$_{\sun}$) & \texttt{Normal} $[1.22,\,0.5]$ \\
            $M_B$ (M$_{\sun}$) & \texttt{Normal} $[1.38,\,0.5]$ \\
            $\pi$  (mas)     & \texttt{Normal} $[22.0,\,0.6]$ \\
            $\gamma_{AB}$ (km s$^{-1}$) & \texttt{Uniform}$[0,\,20]$\\
            &  \\
            \hline
            \hline 
                \end{tabular}
                \label{tab:appendix_priorOuter}
        \end{table}

   \begin{figure*}[]
   \centering
   \includegraphics[width=\textwidth]{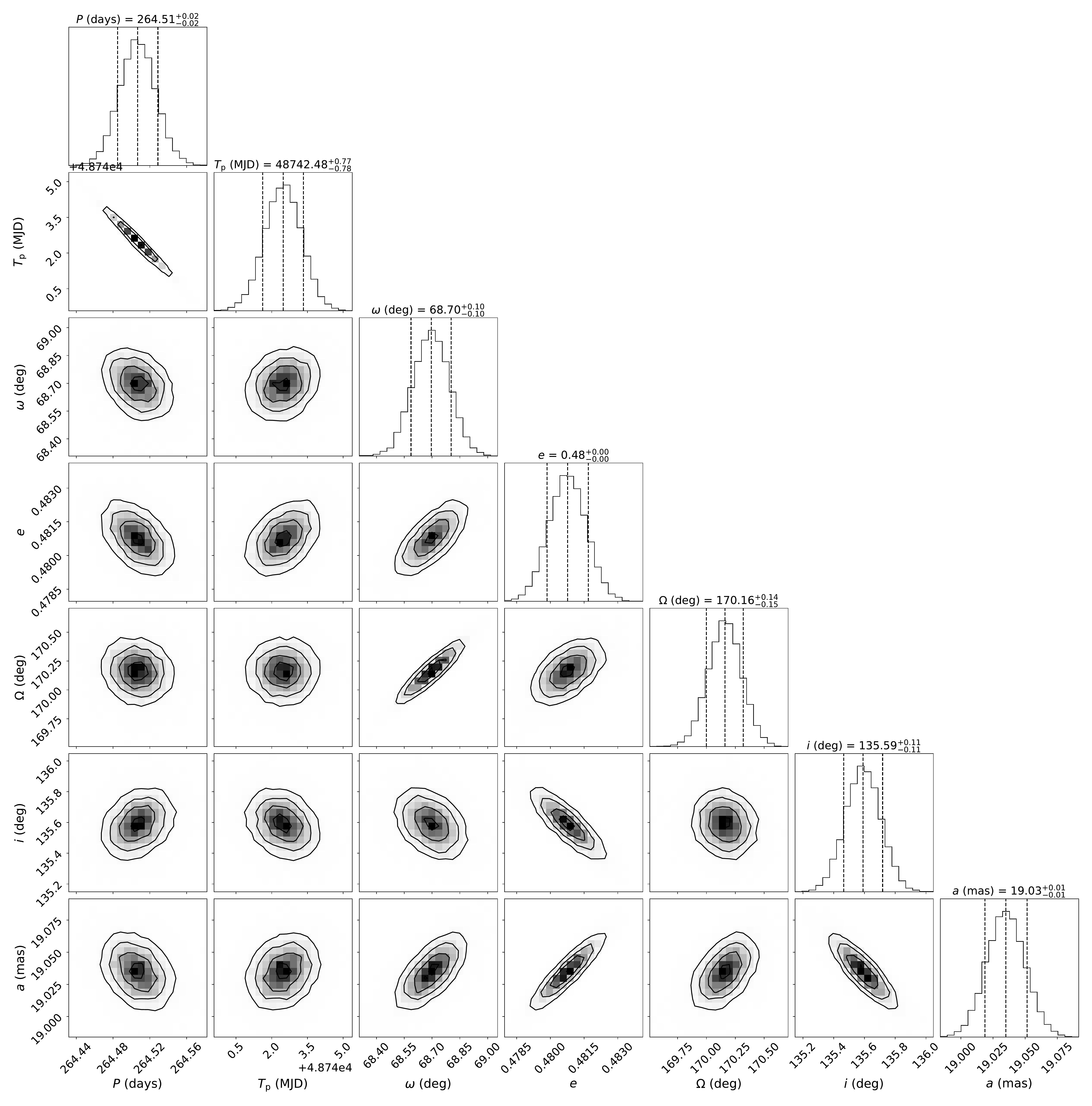}
   \caption{Posterior samples of AaAb orbital parameters. Contoured sub-panels show the distribution of points from the MCMC chains, where high-density regions are indicated by the greyscale and contours. Histogram sub-panels show the posterior distributions, with median and 68\% confidence intervals marked by dashed lines, with titles quantifying those ranges.}
  \label{Fig:AaAb_corner}%
    \end{figure*}
    
   \begin{figure*}[]
   \centering
   \includegraphics[width=\textwidth]{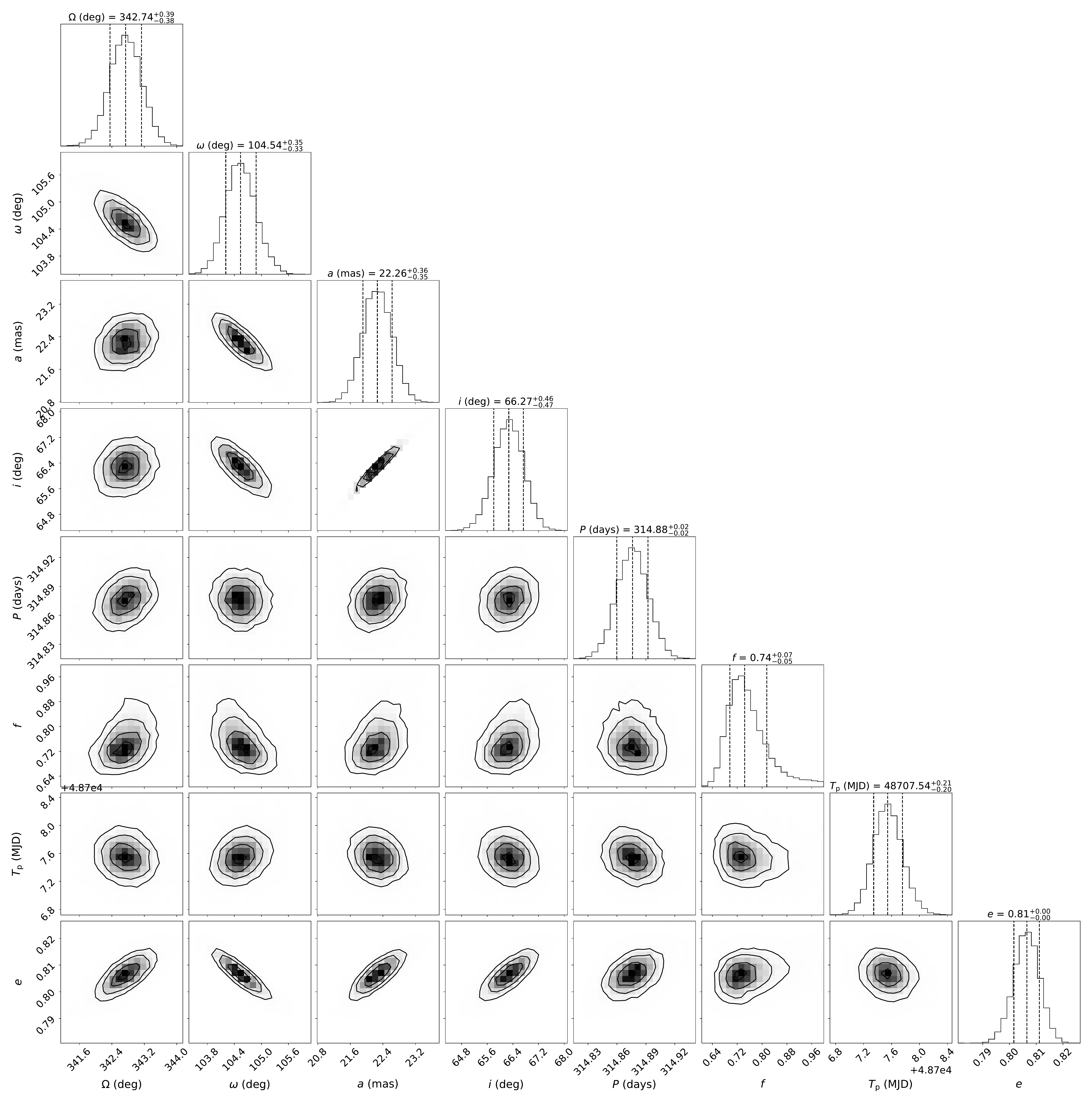}
   \caption{Posterior samples of BaBb orbital parameters. Contoured sub-panels show the distribution of points from the MCMC chains, where high-density regions are indicated by the greyscale and contours. Histogram sub-panels show the posterior distributions, with median and 68\% confidence intervals marked by dashed lines, with titles quantifying those ranges.}
  \label{Fig:BaBb_corner}%
    \end{figure*}

   \begin{figure*}[]
   \centering
   \includegraphics[width=\textwidth]{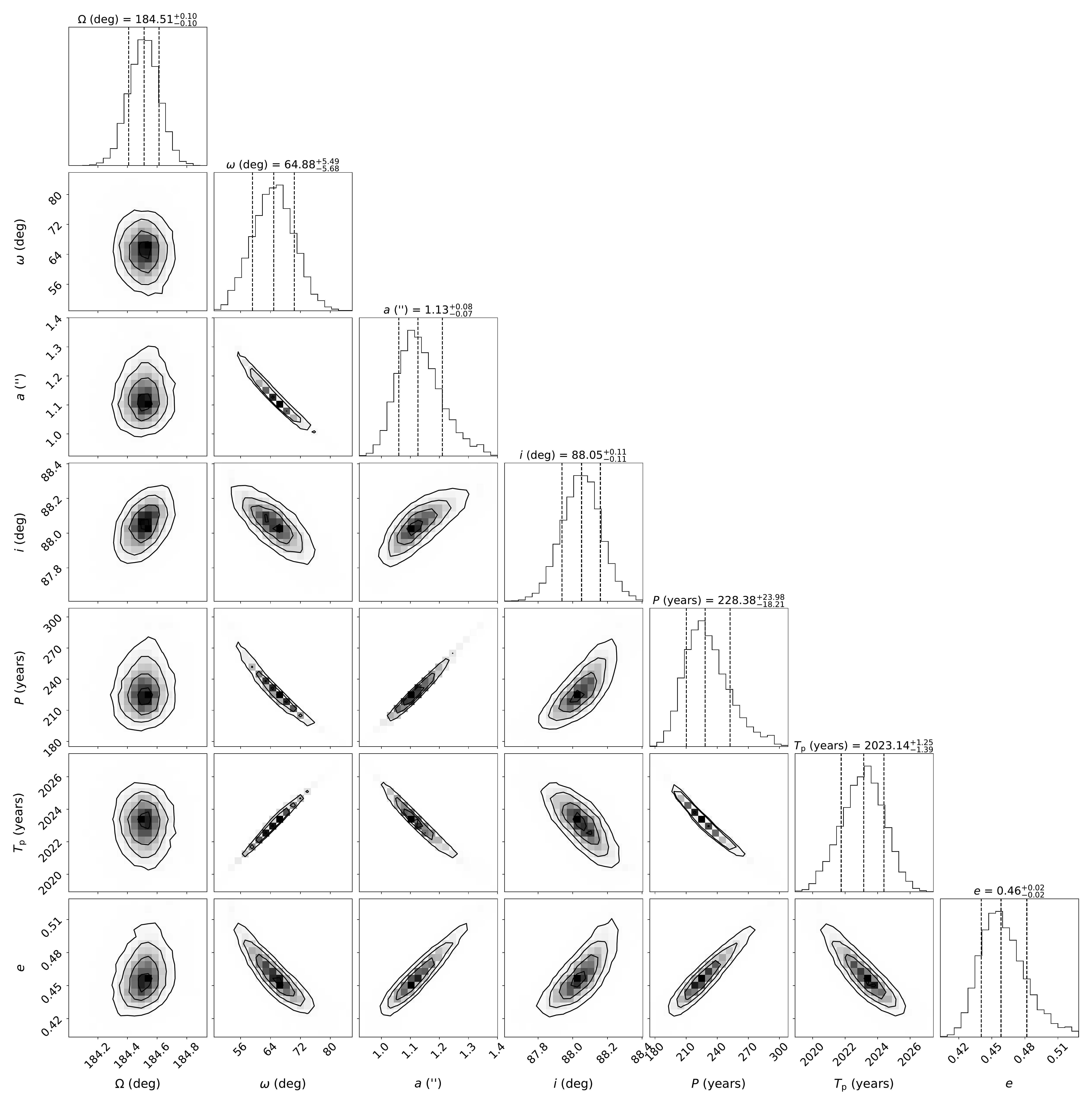}
   \caption{Posterior samples of AB orbital parameters for solution I. Contoured sub-panels show the distribution of points from the MCMC chains, where high-density regions are indicated by the greyscale and contours. Histogram sub-panels show the posterior distributions, with median and 68\% confidence intervals marked by dashed lines, with titles quantifying those ranges.}
  \label{Fig:AB_corner_SolI}%
    \end{figure*}

   \begin{figure*}[]
   \centering
   \includegraphics[width=\textwidth]{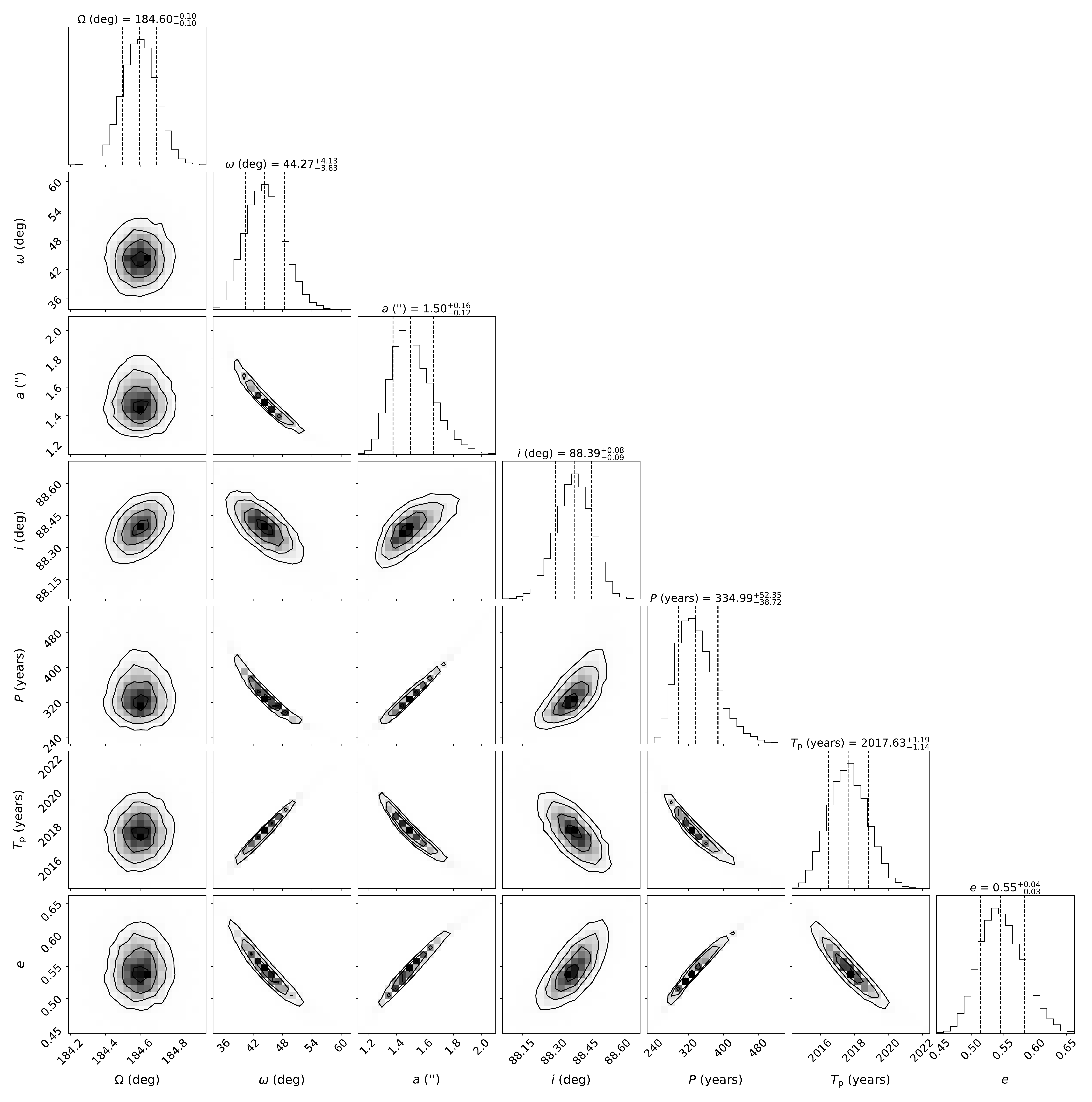}
   \caption{Posterior samples of AB orbital parameters for solution II. Contoured sub-panels show the distribution of points from the MCMC chains, where high-density regions are indicated by the greyscale and contours. Histogram sub-panels show the posterior distributions, with median and 68\% confidence intervals marked by dashed lines, with titles quantifying those ranges.}
  \label{Fig:AB_corner_SolII}%
    \end{figure*}

        \section{N-body simulations}\label{sec:apendix_nbody}

In \texttt{REBOUND}, particles (in that case, stars) are added sequentially to the simulation. Even though a `primary' keyword can be provided to indicate, for instance, that star \#4 is orbiting star \#3, the orbital parameters of the AB orbit are obtained with respect to the centres of mass of AaAb and BaBb, respectively. Therefore, to initialise the simulation, we determined the initial conditions of the four stars. We first added Ba as our heliocentric reference frame, then added Bb by specifying its orbital parameters with respect to Ba and shifted the reference system to the centre of mass of BaBb. Later, we used the AB orbital parameters to simulate a third body with a combined mass $M_\mathrm{Aa}+M_\mathrm{Ab}$ which corresponds to the centre of mass of the A system. We then saved the initial 3D positions $\vec{x}_0$ and velocities $\vec{v}_0$ of this third body `AaAb' using the centre of mass of BaBb as the reference frame. We then set up a new simulation, only for the AaAb system to get the initial positions of Aa and Ab, $\vec{x}_{\mathrm{Aa}, 0}$, $\vec{x}_{\mathrm{Ab}, 0}$ and velocities  $\vec{v}_{\mathrm{Aa}, 0}$,  $\vec{v}_{\mathrm{Ab}, 0}$ with respect to the centre of mass of the AaAb pair. All the positions and velocities for all four stars were calculated at the same reference time, in our case we used $T_0$ of the AB orbit. Finally, we set up the final simulation by adding Ba, followed by Bb by specifying its orbital parameters with respect to Ba, which moved to the centre of mass of BaBb. We then added Aa by specifying its initial position and velocity calculated earlier, the position and velocity are $\vec{x}_0 + \vec{x}_{\mathrm{Aa}, 0}$ and $\vec{v}_0 + \vec{v}_{\mathrm{Aa}, 0}$, respectively, and we then did the same for Ab.

\begin{figure*}
\centering
\includegraphics[width=0.94\textwidth]{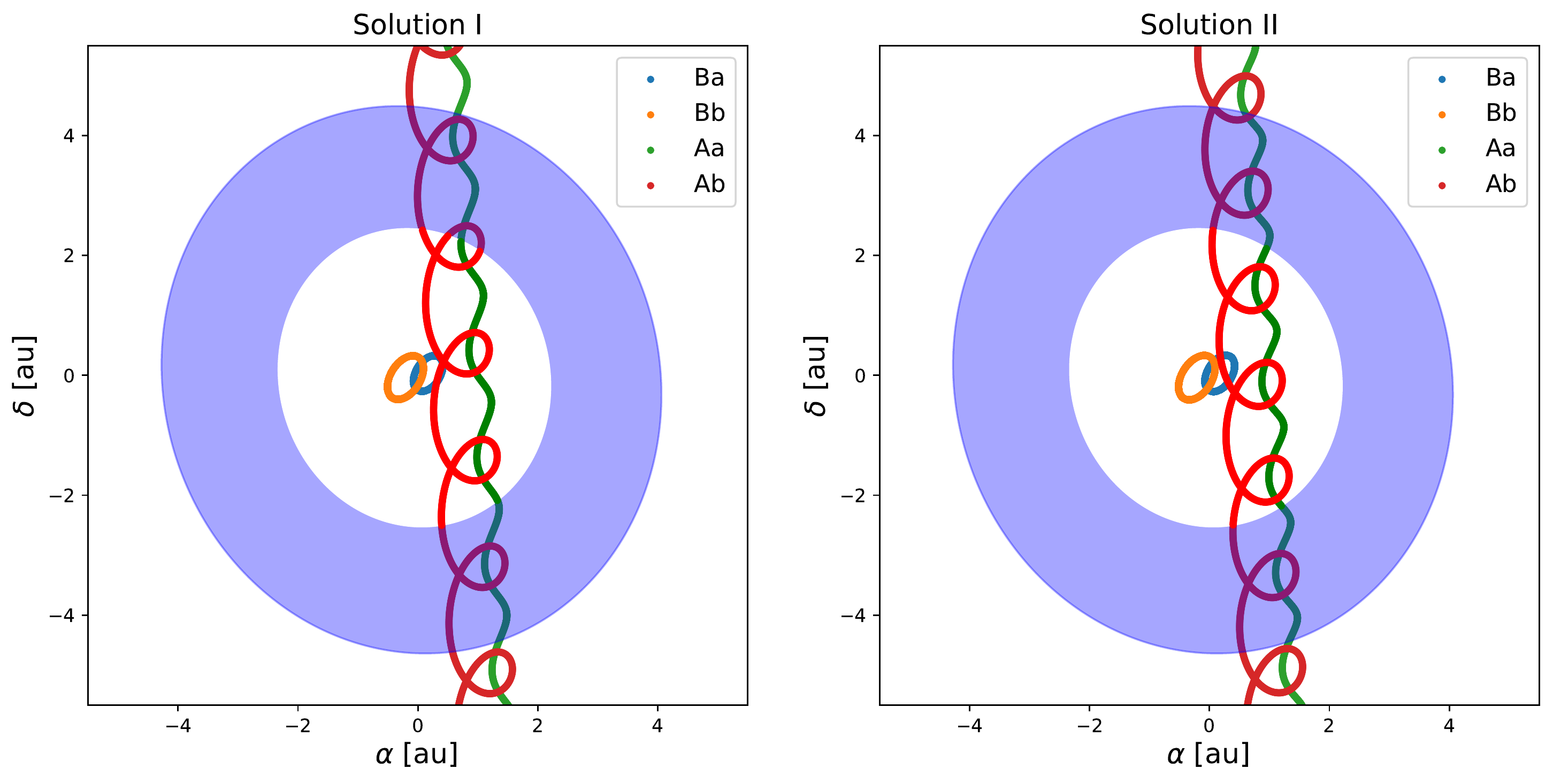}
\caption{Integrated orbits at the times of transit of AaAb behind the disk surrounding BaBb, using the best-fit parameters. The disk and the four orbits are referred to the centre of mass of BaBb located at (0,0).}
\label{Fig:TransitOrbit}
\end{figure*}

Figure\,\ref{Fig:TransitOrbit} shows the positions of the four stars as we integrated the simulation in time for both solutions I and II, overlapped with the location of the disk. The centre of mass of BaBb is located at (0,0).


\section{Flux ratio estimation}\label{sec:appendix_fratio}

We used evolutionary track from \cite{Baraffe2015}, assuming an age of 10 Myr, and synthetic photometry with a BT-Settl model grid, provided by the Spanish Virtual Observatory (SVO) web service \footnote{\url{http://svo2.cab.inta-csic.es/theory/vosa/}} to estimate the flux ratio corresponding to the dynamical masses obtained in this work. The theoretical flux from the BT-Settl model was scaled by the multiplicative dilution factor $M_d = (R/D)^2$, $R$ being the stellar radius and $D$ the distance to the observer (see Tables \ref{tab:flux_ratio_H} and \ref{tab:flux_ratio_V}). 

\begin{table}[h]
\tiny
                \centering
                \caption{Stellar parameters used for the flux ratio estimation in H-band ($1.50 -1.80\,\mu m$) }
                \begin{tabular}{lccccc}
                        \hline
                        \hline  \\
                         & \multicolumn{3}{c}{Adopted stellar parameters} & \multicolumn{2}{c}{predicted observed flux}\\
                         \cline{2-4} \cline{5-6}\\ 
                        Star & $T_{\mathrm{eff}} $ & $log\,g$ & $R$ &  ($\mathrm{[M/H]}=0$)\tablefootmark{a} &  ($\mathrm{[M/H]}=-0.5$)\tablefootmark{b}\\
                        & (K) & & ($R_{\odot}$) & ($erg/cm^2/s/A$) & ($erg/cm^2/s/A$) \\
                          \hline\\
                        Aa  & 4400 & 4.5  & 1.133 & $2.2423\times 10^{-13}$ & $2.1823\times 10^{-13}$ \\
            Ab & 3400 & 4.5 & 0.662 & $3.5559\times 10^{-14}$ & $3.2881\times 10^{-14}$ \\
            & && & & \\
            \hline
            \hline 
                \end{tabular}
                \tablefoot{\tablefoottext{a}{From theoretical flux obtained with the BT-Settl (CIFIST) model \citep{BT-Settl_Allard2013,Caffau2011} multiplied by the dilution factor $M_d$.}\tablefoottext{b}{Same as (a), but using the theoretical flux obtained with theBT-Settl (AGSS2009) model \citep{BT-Settl_Allard2013,Asplund2009}.}}
                \label{tab:flux_ratio_H}
        \end{table} 

\begin{table}[h]
\tiny
                \centering
                \caption{Stellar parameters used for the flux ratio estimation in visible-band ($6040.35-6128.93\,\angstrom$) }
                \begin{tabular}{lccccc}
                        \hline
                        \hline  \\
                        & \multicolumn{3}{c}{Adopted stellar parameters} & \multicolumn{2}{c}{predicted observed flux}\\
                         \cline{2-4} \cline{5-6}\\ 
                        Star & $T_{\mathrm{eff}} $ & $log\,g$ & $R$ &  ($\mathrm{[M/H]}=0$)\tablefootmark{a} &  ($\mathrm{[M/H]}=-0.5$)\tablefootmark{b}\\
                        & (K) & & ($R_{\odot}$) & ($erg/cm^2/s/A$) & ($erg/cm^2/s/A$) \\
                          \hline\\
                        Ba  & 4000 & 4.5  & 1.064 & $2.8660\times 10^{-13}$ & $3.0026\times 10^{-13}$\\
            Bb & 3700 & 4.5 & 0.942 & $1.2269\times 10^{-13}$ & $1.3758\times 10^{-13}$ \\
            & &  &&&\\
            \hline
            \hline 
                \end{tabular}
                \tablefoot{\tablefoottext{a}{From theoretical flux obtained with the BT-Settl (CIFIST) model \citep{BT-Settl_Allard2013,Caffau2011} multiplied by the dilution factor $M_d$.}\tablefoottext{b}{Same as (a), but using the theoretical flux obtained with the BT-Settl (AGSS2009) model \citep{BT-Settl_Allard2013,Asplund2009}.}}
                \label{tab:flux_ratio_V}
        \end{table}

\end{appendix}

\end{document}